\documentclass[superscriptaddress,aps,amsfonts,notitlepage,nofootinbib]{revtex4-1}
\usepackage{xcolor,graphicx}
\usepackage{breqn,amsmath,amssymb}

\usepackage{enumerate}

\usepackage{footmisc}

\usepackage{booktabs}
\usepackage{makecell}

\usepackage{comment}

\usepackage{subcaption}

\usepackage[T1]{fontenc}
\usepackage{natbib,hyperref}

\usepackage{mathtools}

\usepackage{centernot}

\makeatletter
\let\cat@comma@active\@empty
\makeatother

\usepackage{IEEEtrantools}

\begin{document}

\bstctlcite{IEEEexample:BSTcontrol}

\title{Quartic Horndeski-Cartan theories in a FLRW universe}

\author{S. Mironov}
\email{sa.mironov\_1@physics.msu.ru}
\affiliation{Institute for Nuclear Research of the Russian Academy of Sciences, 
60th October Anniversary Prospect, 7a, 117312 Moscow, Russia}
\affiliation{Institute for Theoretical and Mathematical Physics,
MSU, 119991 Moscow, Russia}
\affiliation{NRC, "Kurchatov Institute", 123182, Moscow, Russia}

\author{M. Valencia-Villegas}
\email{mvalenciavillegas@itmp.msu.ru}
\affiliation{Institute for Theoretical and Mathematical Physics,
MSU, 119991 Moscow, Russia}

\begin{abstract}
We consider the Quartic Horndeski theory with torsion on a FLRW background in the second order formalism. We show that there is a one parameter family of Quartic Horndeski Cartan Lagrangians and all such theories only modify the dispersion relations of the graviton and the scalar perturbation that are usually found in the standard Horndeski theory on a torsionless spacetime. In other words, for the theories in this class torsion does not induce new degrees of freedom but it only modifies the propagation. This holds for first order perturbations in spite of a kinetic mixing between the Horndeski scalar with the torsion field in the action. We also show that for most Lagrangians within the family of Quartic Horndeski Cartan theories the dispersion relation of the scalar mode is radically modified. We find only one theory within the family whose scalar mode has a regular wave-like dispersion relation. 
\end{abstract}

\maketitle

\section{Introduction}
Many theoretical developments in modified gravity have been performed in relation to Horndeski theory \cite{creminelli2006starting, Kobayashi:2010cm, Creminelli:2010ba,Kobayashi:2011nu, Easson:2011zy,Luty:2003vm, Libanov:2016kfc,Kobayashi:2016xpl,Volkova:2019jlj,Mironov:2019mye, Mironov:2022quk,Mironov:2020pqh,Mironov:2020mfo} which is a modification of Gravity with higher derivatives in the action, but with second order equations of motion \cite{horndeski1974second,nicolis2009galileon, deffayet2009covariant,Deffayet:2010qz,Padilla:2012dx,Fairlie:1991qe,kobayashi2019horndeski,arai2022cosmological}. These advances have been in part motivated by the possibility to violate the null energy condition in a stable way \cite{rubakov2014null} and to construct interesting cosmological solutions. Despite the advances, the recent measurement of the speed of gravitational waves that  has already constrained many of these theories  makes clear the need to exhaust their broad range of phenomenology \cite{arai2022cosmological}. For instance, Horndeski theory in a different formalism such as metric affine gravity and Palatini, and with torsion and non-metricity have been recently analyzed and it has been pointed out  that new degrees of freedom beyond the usual Horndeski scalar and tensor modes may arise \cite{aoki2018galileon,arai2022cosmological,kubota2021cosmological,helpin2020varying,helpin2020metric,dong2022constraining,davydov2018comparing,dong2022polarization,capozziello2023ghost,dialektopoulos2022classification,bernardo2021well,bernardo2021well2,bahamonde2020post,bahamonde2019can}. 

In this work we start the analysis of Horndeski theories with torsion in the second order (metric) formalism and examine if the degrees of freedom are modified with respect to Horndeski theories on a torsionless spacetime. In the {\it second order formalism} we consider a connection that is {\it a priori} metric compatible and written in terms of the (metric-dependent) Levi-Civita connection and the torsion tensor. One compelling feature in this formalism is that in contrast to Einstein-Cartan theory, torsion may be dynamical because it couples to second derivatives of the scalar in the Lagrangian and simultaneously, the absence of the Ostrogradsky ghost is guaranteed by the Horndeski construction. The latter may not follow for Horndeski theories in other formalisms different than the second order \cite{helpin2020varying, kubota2021cosmological}.

We first show that there is a one parameter family of Quartic Horndeski theories with torsion which reduces to the usual Horndeski theory on a torsionless spacetime. We find in a perturbative expansion at linear order that contrary to the expectation, these Quartic Horndeski Cartan theories on a FLRW background do not introduce additional degrees of freedom and that the torsionful connection only modifies the usual tensor and scalar degrees of freedom that are found in the standard Quartic Horndeski theory without torsion.

We compute the speed of sound for the graviton, which is the same for all the theories with torsion, and find that the subluminality, no ghost and stability conditions are similar to the standard torsionless Horndeski theory. We also show that in most of the Quartic Horndeski Cartan theories the dispersion relation of the scalar mode is radically modified and has no counterpart with the usual scalar mode in the torsionless Horndeski theory. We consider a particular example for the latter and observe that the unusual dispersion relation does not necessarily imply an instability. Furthermore, we show that there are Horndeski Cartan theories within the family for which both the graviton and the scalar mode are simultaneously ghost-free in the high momentum limit, conditional to the assumption that the graviton is also stable and subluminal.  Finally, we find that there is {\it only one} theory within the family in which the scalar field perturbation propagates with a regular wave-like dispersion relation.

We proceed as follows: In section \ref{sec intro} we describe the theories to be analyzed in this work. In section \ref{sec intro torsion} we explicitly introduce torsion in the second order formalism and in section \ref{sec notation} we write the decomposition of the torsion perturbations into irreducible components under rotation group. 

The main results are presented in section \ref{sec main}. In particular, in section \ref{sec classification} we show a classification of the scalar mode according to the parameter of the theory. In section \ref{sec example} we consider a particular example in order to examine the unusual dispersion relation in a subclass of the theories considered. 

We give a summary in section \ref{sec conclusions}. In section \ref{sec discussions} we make an observation on the possibility of strong coupling for cosmological applications of these novel theories and we discuss further avenues of research to assess this question.

\section{Quartic Horndeski Cartan Lagrangians}\label{sec intro}

Lifting the assumption of a Christoffel connection to define covariant differentiation on the spacetime introduces new fields in the theory. In this work we assume from the beginning, when we formulate the theory, that the connection can be expressed in terms of the metric and a torsion tensor. Furthermore, we assume a vanishing nonmetricity. Namely, we only consider metric compatible covariant derivatives. 

This approach is usually known as second order formalism, in contrast to previous works where one could, for instance, start from a connection initially assumed to be independent of the metric \cite{aoki2018galileon, arai2022cosmological,kubota2021cosmological,helpin2020varying}.

Within the second order formalism the natural approach to Horndeski theory on a spacetime with torsion is to promote torsionless to torsionful covariant derivatives in the Horndeski Lagrangian. However, this prescription "Torsionless to Torsionful" does not lead to an unique choice of Lagrangian function. More precisely, there is at least one parameter family of Lagrangian functions which reduces to the usual Quartic Horndeski  Lagrangian when we assume a  Christoffel connection.

To write this precisely let us start with the Quartic Horndeski Lagrangian in the generalized Galileon notation on a spacetime without torsion
\begin{eqnarray}
\mathcal{L}_4=G_4(\phi,X){R}+G_{4,X}\left(\left({\nabla}_\mu{\nabla}^\mu\phi\right)^2-\left({\nabla}_\mu{\nabla}_\nu\phi\right)^2 \right)\, ,\label{eqn G4lag}
\end{eqnarray}
where $G_4$ is an arbitrary function of  $\phi$ and $X=-\frac{1}{2}g^{\mu\nu}\partial_\mu\phi \partial_\nu\phi$, $g$ is the metric with mostly $+$ signature, $G_{4,X}=\partial G_4/\partial X$, and the metric compatible covariant derivative on a vector $V$ on a spacetime without torsion is written as
\begin{eqnarray}
{\nabla}_\mu V^\nu=\partial_\mu V^\nu+{\Gamma}^{\nu}_{\mu\lambda}V^{\lambda}\,\, \label{eqn torsionless covd}
\end{eqnarray}
where
\begin{eqnarray}
\Gamma^{\rho}_{\mu\nu}&=&\frac{1}{2}g^{\rho\sigma}\left(\partial_\mu g_{\nu\sigma}+\partial_\nu g_{\mu\sigma}-\partial_\sigma g_{\mu\nu}\right)\,,
\end{eqnarray}
such that
\begin{eqnarray}
\Gamma^{\rho}_{\mu\nu}&=& \Gamma^{\rho}_{\nu\mu}\,. \label{eqn symm connection}
\end{eqnarray}
As such, it is clear that two torsionless covariant derivatives commute on the scalar
\begin{equation}
\left[{\nabla}_\mu,{\nabla}_\nu\right]\phi=0 \,,
\end{equation}
and therefore there is no ambiguity in the contraction of Lorentz indices in the rightmost term $G_{4,X}\,\left({\nabla}_\mu{\nabla}_\nu\phi\right)^2 $ in the Quartic Horndeski Lagrangian without torsion (\ref{eqn G4lag}).\bigskip

Now we consider the case on a spacetime with torsion such that the connection on the spacetime is not symmetric (we use tilde notation for torsionful quantities)
\begin{eqnarray}
\tilde{\Gamma}^{\nu}_{\mu\lambda}\neq \tilde{\Gamma}^{\nu}_{\lambda \mu}\,.\label{eqn basicassumption}
\end{eqnarray}
Writing the torsionful metric compatible covariant derivative as (we stick to the convention to sum over the second index of the non symmetric connection)
\begin{eqnarray}
\tilde{\nabla}_\mu V^\nu=\partial_\mu V^\nu+\tilde{\Gamma}^{\nu}_{\mu\lambda}V^{\lambda}\,,\label{eqn covd}
\end{eqnarray}
we can write a one parameter ($c$) family of Quartic Horndeski Cartan Lagrangians by considering all possible contractions with the metric of the terms of the form $(\tilde{\nabla}_\mu\tilde{\nabla}_\nu\phi)^2$ in the $G_{4,X}$ "counterterm", where Torsion only appears implicitly in the torsionful covariant derivatives
\begin{eqnarray}
G_{4,X} \left(\left(\tilde{\nabla}_\mu\tilde{\nabla}^\mu\phi\right)^2\,+\,c\, \left(\tilde{\nabla}_\mu\tilde{\nabla}_\nu\phi\right) \tilde{\nabla}^\mu\tilde{\nabla}^\nu\phi \,+\, s\, \left(\tilde{\nabla}_\mu\tilde{\nabla}_\nu\phi\right) \tilde{\nabla}^\nu\tilde{\nabla}^\mu\phi\right)\,, \label{eqn L cases}
\end{eqnarray}
with $c+s=-1$, such that the terms (\ref{eqn L cases}) reduce to the standard counterterm proportional to $G_{4,X}$ in the Horndeski theory (\ref{eqn G4lag}), when we assume a Christoffel connection (namely, by the prescription $\tilde{\Gamma}\rightarrow {\Gamma}$). Thus, these Horndeski Cartan theories take the form
\begin{eqnarray}
\mathcal{S}_{4c}&=&\int \,\text{d}^4x\,\sqrt{-g}\, \mathcal{L}_{4c}\, ,\label{eqn G4Tlag}
\end{eqnarray}
\begin{eqnarray}
\mathcal{L}_{4c}&=&\,G_4(\phi,X)\tilde{R}+G_{4,X}\left(\left(\tilde{\nabla}_\mu\tilde{\nabla}^\mu\phi\right)^2-\left(\tilde{\nabla}_\mu\tilde{\nabla}_\nu\phi\right) \tilde{\nabla}^\nu\tilde{\nabla}^\mu\phi -c\,\left(\tilde{\nabla}_\mu\tilde{\nabla}_\nu\phi\right) \left[\tilde{\nabla}^\mu,\tilde{\nabla}^\nu\right]\phi \right)\,,
\end{eqnarray}

where $\tilde{R}$ is the Ricci scalar with torsion and $c$ is a real constant. Below we will show that the choice of parameter $c$ is important to determine the dynamics. Furthermore, let us stress that the term with coupling $c$ does not introduce higher derivatives because the antisymmetric two tensor $c\,\left[\tilde{\nabla}^\mu,\tilde{\nabla}^\nu\right]\phi $ clearly vanishes the symmetric second order derivatives on the scalar $\partial_\mu\partial_\nu \phi$.

Indeed, let us note from the start that there are no higher than second order derivatives of any of the fields in the equations of motion derived from (\ref{eqn G4Tlag}) for all values of $c$ (See the Appendix \ref{asec eqns}). Hence, as expected from Horndeski theories in the second order formalism, all the Lagrangians in (\ref{eqn G4Tlag}) are free of the Ostrogradsky ghost.

\subsection{Torsion in the Quartic Horndeski Lagrangian in the second order formalism}\label{sec intro torsion}

As opposed to the usual Horndeski $G_4$ Lagrangian where the metric and the scalar are the only fields, now there is an additional Torsion field that is necessary to specify the geometry of the spacetime \cite{shapiro2002physical}.

To write explicitly the torsion in the Quartic Horndeski Cartan action (\ref{eqn G4Tlag}) in the second order formalism (See for instance \cite{shapiro2002physical,arai2022cosmological}), we proceed as follows: provided the assumption (\ref{eqn basicassumption}) and the fact that every difference of connections is a tensor, we define the torsion tensor as
\begin{eqnarray}
T^{\rho}{}_{\mu\nu}=\tilde{\Gamma}^{\rho}_{\mu\nu}-\tilde{\Gamma}^{\rho}_{\nu\mu}\,,
\end{eqnarray}
and for latter convenience the contortion tensor as
\begin{eqnarray}
K^{\rho}{}_{\mu\nu}&=&-\frac{1}{2}\left(T_{\nu}{}^{\rho}{}_{\mu}+T_{\mu}{}^{\rho}{}_{\nu}+T^{\rho}{}_{\mu\nu}\right)\,,\label{eqn ktrelation}
\end{eqnarray}
where we notice the antisymmetry
\begin{eqnarray}
T^{\rho}{}_{\mu\nu}&=& -T^{\rho}{}_{\nu\mu}\\
K_{\mu\nu\sigma}&=& -K_{\sigma\nu\mu}\,.\label{eqn ksymm}
\end{eqnarray}

Now, a direct computation shows that the torsionful connection and the Christoffel connection are related by the contortion tensor as
\begin{eqnarray}
\tilde{\Gamma}^{\rho}_{\mu\nu}&=&\Gamma^{\rho}_{\mu\nu}-K^{\rho}{}_{\mu\nu}\,.
\end{eqnarray}

With these definitions for torsion, contortion and previous relations to the Christoffel connection, we can rewrite the torsionful covariant derivative explictly in terms of contortion and the covariant derivative ($\nabla$) associated with the Christoffel symbol as
\begin{eqnarray}
\tilde{\nabla}_\mu V^\nu=\nabla_\mu V^\nu-K^{\nu}{}_{\mu\lambda}V^{\lambda}\,,\label{eqn covds}
\end{eqnarray}
and we can write the commutator of torsionful covariant derivatives on a scalar as
\begin{equation}
\left[\tilde{\nabla}_\mu,\tilde{\nabla}_\nu\right]\phi=-T^\lambda{}_{\mu\nu}\, \partial_\lambda \phi\,.
\end{equation}
With this commutator we can rewrite the action (\ref{eqn G4Tlag}) in a form more reminiscent of the usual Quartic Horndeski, but with torsionful covariant derivatives plus a lower derivative $c$ term. This last term parameterizes different choices of Lorentz index contractions in the Quartic Horndeski Cartan theories. In terms of contortion, (\ref{eqn G4Tlag}) takes the form
\begin{eqnarray}
\mathcal{S}_{4c}&=& \int \,\text{d}^4x\,\sqrt{-g}\,\Big(\, G_4(\phi,X)\tilde{R}+G_{4,X}\left(\left(\tilde{\nabla}_\mu\tilde{\nabla}^\mu\phi\right)^2-\left(\tilde{\nabla}_\mu\tilde{\nabla}_\nu\phi\right) \tilde{\nabla}^\mu\tilde{\nabla}^\nu\phi \right)\nonumber\\
&+&(1-c)G_{4,X}K_{\nu\mu\sigma}\left(K^{\nu\mu\lambda}-K^{\mu\nu\lambda}\right)\tilde{\nabla}_\lambda\phi \tilde{\nabla} ^\sigma \phi\Big)\,.
\end{eqnarray}

Finally, let us write $\tilde{R}$ in terms of the Ricci scalar without torsion ($R$) and contortion, as,
\begin{eqnarray}
\tilde{R}=R+K_{\mu\rho\nu} K^{\mu\nu\rho} + K^{\mu}{}_{\mu}{}^{\nu} K_{\nu}{}^{\rho}{}_{\rho}  + 2 \nabla_{\nu}K^{\mu}{}_{\mu}{}^{\nu}\,.
\end{eqnarray}
All in all, with the previous definitions we can rewrite the Quartic Horndeski theories (\ref{eqn G4Tlag}) in terms of three explicit tensor fields: namely, with the metric, the scalar and the contortion tensor $K$ as the three fundamental fields\footnote{Let us notice that instead of $K$, we could have chosen the torsion tensor $T$ as fundamental,  by means of equation (\ref{eqn ktrelation}).}
\begin{eqnarray}
\mathcal{S}_{4c}&=& \int \,\text{d}^4x\,\sqrt{-g}\,\Big(\, G_4 (R+K_{\mu\rho\nu} K^{\mu\nu\rho} + K^{\mu}{}_{\mu}{}^{\nu} K_{\nu}{}^{\rho}{}_{\rho}  + 2 \nabla_{\nu}K^{\mu}{}_{\mu}{}^{\nu}) +G_{4X}(\nabla_{\nu}\nabla^{\nu}\phi + K^{\rho\nu}{}_{\nu} \nabla_{\rho}\phi)^2 \nonumber\\
&-&  G_{4X} (K^{\rho}{}_{\gamma\mu} \nabla_{\rho}\phi + \nabla_{\gamma}\nabla_{\mu}\phi) (K^{\nu\gamma\mu} \nabla_{\nu}\phi + \nabla^{\gamma}\nabla^{\mu}\phi)\nonumber \\
&+&(1-c)G_{4,X}K_{\nu\mu\sigma}\left(K^{\nu\mu\lambda}-K^{\mu\nu\lambda}\right){\nabla}_\lambda\phi {\nabla} ^\sigma \phi\Big)\,. \label{eqn G4Texplicitlag}
\end{eqnarray} 

\subsection{Linearization: Decomposition of contortion perturbations into irreducible components}\label{sec notation}
We will perform a perturbative expansion at linear order about a spatially flat FLRW background. It is convenient to decompose the perturbations into irreducible components under small rotation group as follows: we consider the perturbed metric
\begin{eqnarray}
\textrm{d}s^2=\left(\eta_{\mu\nu}+\delta g_{\mu\nu}\right)\textrm{d}x^\mu\, \textrm{d}x^\nu
\end{eqnarray}
where 
\begin{eqnarray}
\eta_{\mu\nu}\textrm{d}x^\mu\, \textrm{d}x^\nu= a^2(\eta)\left(-\textrm{d}\eta^2+\delta_{ij}\, \textrm{d}x^i \,\textrm{d}x^j \right)\label{eqn backgroundmetric}
\end{eqnarray}
is a spatially flat FLRW background metric, $\eta$ is conformal time, and we denote spatial indices with latin letters such as $i=1,2,3$ and space-time indices with greek letters, such as $\mu=0,1,2,3$. The metric perturbation is written as 
\begin{eqnarray}
\delta g_{\mu\nu}\,\textrm{d}x^\mu\, \textrm{d}x^\nu
=a^2(\eta)\left(-2\,\alpha\,\textrm{d}\eta^2+2\left(\partial_i B+S_i\right) \textrm{d}\eta \, \textrm{d}x^i+\left(-2\,\psi\, \delta_{ij}+2\,\partial_i\partial_j E+\partial_i F_j+\partial_j F_i+2\,h_{ij}\right) \textrm{d}x^i \, \textrm{d}x^j \right)\,,
\end{eqnarray}
with $\alpha,\, B,\, \psi,\, E$  scalar perturbations, $S_i,\, F_i$  transverse vector perturbations, and $h_{ij}$, a symmetric, traceless and transverse tensor perturbation.\bigskip

For the contortion perturbation, which satisfies the symmetry (\ref{eqn ksymm}), there are $24$ independent components that can be written in terms of irreducible components under small rotation group as: eight scalars denoted as $C^{\scalebox{0.5}{(n)}} $ with $n=1, \dots , 8$, six (two-component) transverse vectors denoted as $V^{\scalebox{0.5}{(m)}}_i$ with $m=1,\dots, 6$ and two (two-component) traceless, symmetric, transverse tensors $T^{\scalebox{0.5}{(1)}}_{ij},\, T^{\scalebox{0.5}{(2)}}_{ij}$. 

Explicitly, the decomposition of contortion perturbation reads, for the scalar sector
\begin{eqnarray}
\delta K^{\text{scalar}}_{i00}&=&\partial_i C^{\scalebox{0.5}{(1)}} \nonumber \\
\delta K^{\text{scalar}}_{ij0}&=& \partial_i \partial_j C^{\scalebox{0.5}{(2)}} +\delta_{ij} C^{\scalebox{0.5}{(3)}} +\epsilon_{ijk} \partial_k C^{\scalebox{0.5}{(4)}} \nonumber \\
\delta K^{\text{scalar}}_{i0k}&=&\epsilon_{ikj} \partial_j C^{\scalebox{0.5}{(5)}} \nonumber \\
\delta K^{\text{scalar}}_{ijk}&=&\left(\delta_{ij} \partial_k-\delta_{kj} \partial_i\right) C^{\scalebox{0.5}{(6)}} +\epsilon_{ikl} \partial_l \partial_j C^{\scalebox{0.5}{(7)}} +\left(\epsilon_{ijl} \partial_l \partial_k-\epsilon_{kjl} \partial_l \partial_i\right) C^{\scalebox{0.5}{(8)}} \label{eqn kspert}\,,
\end{eqnarray}
for the vector sector
\begin{eqnarray}
\delta K^{\text{vector}}_{i00}&=& V^{\scalebox{0.5}{(1)}}_i\nonumber\\
\delta K^{\text{vector}}_{ij0}&=& \partial_i V^{\scalebox{0.5}{(2)}}_j+ \partial_j V^{\scalebox{0.5}{(3)}}_i\nonumber\\
\delta K^{\text{vector}}_{i0k}&=& \partial_i V^{\scalebox{0.5}{(4)}}_k- \partial_k V^{\scalebox{0.5}{(4)}}_i\nonumber\\
\delta K^{\text{vector}}_{ijk}&=&\delta_{ij} V^{\scalebox{0.5}{(5)}}_k-\delta_{kj} V^{\scalebox{0.5}{(5)}}_i + \partial_j \partial_i V^{\scalebox{0.5}{(6)}}_k-\partial_j \partial_k V^{\scalebox{0.5}{(6)}}_i \label{eqn kvpert}\,,
\end{eqnarray}
and for the tensor sector
\begin{eqnarray}
\delta K^{\text{tensor}}_{ij0}&=& T^{\scalebox{0.5}{(1)}}_{ij}\nonumber\\
\delta K^{\text{tensor}}_{ijk}&=& \partial_i T^{\scalebox{0.5}{(2)}}_{jk}-\partial_k T^{\scalebox{0.5}{(2)}}_{ji} \label{eqn ktpert}\,,
\end{eqnarray}
where we have not written explicitly the vanishing components and those related to (\ref{eqn kspert}-\ref{eqn ktpert}) by the symmetry (\ref{eqn ksymm}). All in all, the components of contortion perturbation are
\begin{eqnarray}
\delta K_{i\mu\nu}= \delta K^{\text{scalar}}_{i\mu\nu}+ \delta K^{\text{vector}}_{i\mu\nu}+ \delta K^{\text{tensor}}_{i\mu\nu}\,.
\end{eqnarray}
On the other hand, the non-vanishing components of the background contortion tensor on a homogeneous and isotropic background spacetime are 
\begin{eqnarray}
{}^{0}K_{0jk}=x(\eta)\delta_{jk}\nonumber \\
{}^{0}K_{ijk}=y(\eta)\epsilon_{ijk}\,,
\end{eqnarray}
such that we write at linearized level the contortion tensor with all indices down as
\begin{eqnarray}
K_{\mu\nu\sigma}= {}^{0}K_{\mu\nu\sigma}+ \delta K_{\mu\nu\sigma}
\end{eqnarray}
Finally, let us write the scalar field as 
\begin{eqnarray}
\phi=\varphi(\eta)+\Pi
\end{eqnarray}
where $\Pi$ is a spacetime dependent scalar field perturbation and $\varphi$ is the background scalar field.

All in all, there are $4$ background quantities for the scalar, metric and contortion: $\varphi,\, a,\, x,\, y$ which satisfy $5$ equations of motion, of which only $4$ are independent. Namely, first using the equation for ${}^{0}K_{ijk}$
\begin{eqnarray}
\mathcal{E}_{K_{ijk}} =\epsilon_{ijk}\,\frac{2}{a^6}G_{4}\,y=0\,,
\end{eqnarray}
which fixes $y=0$, we have
\begin{dmath}
{\mathcal{E}}_{g_{00}}= (\,{x} +  \,{a} \,\dot{a})\left(\frac{3 \,{G_{4}} (\,{x} -  \,{a} \,\dot{a})}{\,{a}^8} -  \frac{3 \,{G_{4,{\phi}}} \,\dot{\varphi}}{\,{a}^6} + \frac{6 \,{G_{4,X}}  \,(2\,x+a\,\dot{a})\dot{\varphi}^2}{\,{a}^{10}} -  \frac{3 \,{G_{4,{\phi}X}} \,\dot{\varphi}^3}{\,{a}^8} + \frac{3 \,{G_{4,XX}} (\,{x} + \,{a} \,\dot{a}) \,\dot{\varphi}^4}{\,{a}^{12}}\right)\,,
\end{dmath}
\begin{dmath}
{\mathcal{E}}_{g_{ij}}= \delta_{ij}\,\left(\frac{\,{G_{4}} \Bigl(- \,{x}^2 + \,{a}^4 \bigl(\frac{\,\dot{a}^2}{\,{a}^2} + 2 (\frac{\,\ddot{a}}{\,{a}} -  \frac{\,\dot{a}^2}{\,{a}^2})\bigr)\Bigr)}{\,{a}^8} + \frac{\,{G_{4,{\phi}{\phi}}} \,\dot{\varphi}^2}{\,{a}^4} + \frac{2 \,{G_{4,XX}} (\,{x} + \,{a} \,\dot{a}) \,\dot{\varphi}^3 (- \,\ddot{\varphi} + \frac{\,\dot{a} \,\dot{\varphi}}{\,{a}})}{\,{a}^{10}} + \frac{\,{G_{4,{\phi}X}} \,\dot{\varphi}^2 \bigl(-2 \,{x} \,\dot{\varphi} + \,{a}^2 (\,\ddot{\varphi} -  \frac{3 \,\dot{a} \,\dot{\varphi}}{\,{a}})\bigr)}{\,{a}^8} + \frac{\,{G_{4,{\phi}}} \bigl(- \,{x} \,\dot{\varphi} + \,{a}^2 (\,\ddot{\varphi} + \frac{\,\dot{a} \,\dot{\varphi}}{\,{a}})\bigr)}{\,{a}^6} + \frac{\,{G_{4,X}} \,\dot{\varphi} \Bigl(- \,{x}^2 \,\dot{\varphi} + \,{a}^4 \bigl(- \frac{2 \,\ddot{\varphi} \,\dot{a}}{\,{a}} + \frac{\,\dot{a}^2 \,\dot{\varphi}}{\,{a}^2} - 2 (\frac{\,\ddot{a}}{\,{a}} -  \frac{\,\dot{a}^2}{\,{a}^2}) \,\dot{\varphi}\bigr) + \,{a}^2 (-4 \,{x} \,\ddot{\varphi} + \frac{6 \,{x} \,\dot{a} \,\dot{\varphi}}{\,{a}} - 2 \,\dot{\varphi} \,\dot{x})\Bigr)}{\,{a}^{10}}\right)\,,
\end{dmath}
\begin{dmath}
{\mathcal{E}}_{K_{ij0}}=\delta_{ij}\,\left(\frac{2 \,{G_{4}} \,{x}}{\,{a}^6} -  \frac{\,{G_{4,{\phi}}} \,\dot{\varphi}}{\,{a}^4} + \frac{2 \,{G_{4,X}} (\,{x} + \,{a} \,\dot{a}) \,\dot{\varphi}^2}{\,{a}^8}\right)\,,
\end{dmath}
\begin{eqnarray}
{\mathcal{E}}_{\varphi}=-\frac{2\,a^2}{\dot{\varphi}^2}\,\left(\dot{\mathcal{E}}_{g_{00}}+\left(5\, {\mathcal{E}}_{g_{00}}+3\, {\mathcal{E}}_{g_{ii}}\right)\frac{\dot{a}}{a}\right)+\frac{6\,x}{\dot{\varphi}^2}\left(\dot{\mathcal{E}}_{K_{ii0}}+4\, {\mathcal{E}}_{K_{ii0}}\frac{\dot{a}}{a}\right)\,,\label{eqn eomscalarin}
\end{eqnarray}
where repeated spatial indices are not summed in  expression (\ref{eqn eomscalarin}) and the remaining background equations are
\begin{equation}\begin{array}{cccc}
{\mathcal{E}}_{\varphi}=0 \,,&
{\mathcal{E}}_{g_{00}}=0\,, & {\mathcal{E}}_{g_{ij}}=0\,, & {\mathcal{E}}_{K_{ij0}}=0\,, \label{eqn backgroundeoms}
\end{array}\end{equation}
where we denote derivative with respect to conformal time $\eta$ with dot as, for instance, $\dot{a}=\partial a/\partial \eta$.

\subsubsection{Little gauge transformations}

The action (\ref{eqn G4Texplicitlag}) at quadratic order in perturbations  is invariant under the following gauge transformations of metric perturbations
\begin{align}
\alpha &\rightarrow \alpha -\dot{\xi}^{0}-\frac{\dot{a}}{a} \xi^{0} & B &\rightarrow B +\xi^{0}-\dot{\xi} & \psi &\rightarrow \psi + \frac{\dot{a}}{a} \xi^{0} &  E &\rightarrow E -\xi
\end{align}
\begin{align}
S_i &\rightarrow S_i-\dot{\xi}_i & F_i& \rightarrow F_i - \xi_i & h_{ij}&\rightarrow h_{ij}\,, \nonumber
\end{align}
the transformation of the scalar field perturbation
\begin{equation}
\Pi\rightarrow \Pi -\xi^{0} \dot{\varphi}\,,\label{eqn transfPi}
\end{equation}
and the gauge transformations of torsion perturbations (in momentum space)
\begin{align}
T^{\scalebox{0.5}{(1)}}_{ij}&\rightarrow  T^{\scalebox{0.5}{(1)}}_{ij} & T^{\scalebox{0.5}{(2)}}_{ij}&\rightarrow  T^{\scalebox{0.5}{(2)}}_{ij} & V^{\scalebox{0.5}{(5)}}_{i}&\rightarrow  V^{\scalebox{0.5}{(5)}}_{i} & V^{\scalebox{0.5}{(6)}}_{i}&\rightarrow  V^{\scalebox{0.5}{(6)}}_{i}
\end{align}
\begin{align}
V^{\scalebox{0.5}{(1)}}_{i}&\rightarrow  V^{\scalebox{0.5}{(1)}}_{i}+\dot{\xi}_{i}\, x & V^{\scalebox{0.5}{(2)}}_{i}&\rightarrow  V^{\scalebox{0.5}{(2)}}_{i}+\xi_{i}\, x - \dot{\omega}_i \,y & V^{\scalebox{0.5}{(3)}}_{i}&\rightarrow  V^{\scalebox{0.5}{(3)}}_{i}+\xi_{i}\, x + \dot{\omega}_i\,y & V^{\scalebox{0.5}{(4)}}_{i}&\rightarrow  V^{\scalebox{0.5}{(4)}}_{i}+ \dot{\omega}_i\,y \nonumber
\end{align}
\begin{align}
C^{\scalebox{0.5}{(1)}} & \rightarrow  C^{\scalebox{0.5}{(1)}} + \dot{\xi}\, x & C^{\scalebox{0.5}{(2)}} & \rightarrow  C^{\scalebox{0.5}{(2)}} + 2 \xi \, x & C^{\scalebox{0.5}{(3)}} & \rightarrow  C^{\scalebox{0.5}{(3)}} +\xi^{0}\, \dot{x} + \dot{\xi}^{0}\, x & C^{\scalebox{0.5}{(4)}} & \rightarrow  C^{\scalebox{0.5}{(4)}} -\dot{\xi}\, y \nonumber
\end{align}
\begin{align}
C^{\scalebox{0.5}{(5)}} & \rightarrow  C^{\scalebox{0.5}{(5)}} + \dot{\xi}\, y & C^{\scalebox{0.5}{(6)}} & \rightarrow  C^{\scalebox{0.5}{(6)}} +\xi^{0}\, x & C^{\scalebox{0.5}{(7)}} & \rightarrow  C^{\scalebox{0.5}{(7)}} +\xi\, y - \frac{1}{\vert \vec{p}\vert^2} \dot{y}\xi^0 & C^{\scalebox{0.5}{(8)}} & \rightarrow  C^{\scalebox{0.5}{(8)}} -\xi\, y + \frac{1}{\vert \vec{p}\vert^2} \dot{y}\xi^0 \,,\nonumber
\end{align}
where we have decomposed the gauge $4-$vector into two scalar gauge parameters $\xi^0(\eta,\vec{p})$, $\xi(\eta,\vec{p})$ and the transverse vector $\xi_i(\eta,\vec{p}) $ (with $\partial_i \xi_i=0$). For convenience, we have occasionally written the transverse vector $\xi_i$ in terms of $\omega_i$ as
\begin{align}
\xi_k &=\epsilon_{ijk}\partial_i \omega_j & \partial_i \omega_i &=0\,.
\end{align}   

\section{Linearized Dynamics in Quartic Horndeski Cartan theories}\label{sec main}

In contrast to the case of Einstein-Cartan \cite{shapiro2002physical}, we do not generally expect torsion to decouple as constraint equations in the Quartic Horndeski Cartan theory because there are terms of the form $G_4 \, \tilde\nabla K $ and $G_{4,X}\left(\, \tilde\nabla\phi\right) \left(\, \tilde\nabla \, \tilde\nabla\phi\right)K $ in the action (\ref{eqn G4Texplicitlag}). These terms generate second derivatives of contortion ($K$)  in the Euler Lagrange equation for the scalar ($\phi$) and second derivatives of the scalar in the Euler Lagrange equation for contortion.

More precisely, the Euler Lagrange equations for $\phi,\, g_{\mu\nu} $ and $K^\mu{}_{\nu\sigma} $ computed from the action (\ref{eqn G4Texplicitlag}) are
\begin{equation}
\begin{array}{ccccc}
{\mathcal{E}}_{\phi} (\tilde\nabla^2 K, \,\tilde\nabla^2 \phi,\, \partial^2 g) =0 \,,& \,{} &
{\mathcal{E}}_{g_{\mu\nu}} (\tilde\nabla^2 \phi,\, \partial^2 g) =0\,, & \,{} & {\mathcal{E}}_{K^\mu{}_{\nu\sigma}}(\tilde\nabla^2 \phi) =0\,, \label{eqn fieldeoms}
\end{array}
\end{equation}
where we have shown the dependances of ${\mathcal{E}}_{\phi},\,{\mathcal{E}}_{g_{\mu\nu}},\, {\mathcal{E}}_{K^\mu{}_{\nu\sigma}}$ on the highest derivatives of the fields.

They take the form
\begin{eqnarray}
\mathcal{E}_\phi=2\,G_{4X}\,\partial^\lambda \phi\left(\tilde\nabla_\lambda \,\tilde\nabla_\mu K^{\nu}{}_{\nu}{}^\mu-\,\tilde\nabla_\nu \,\tilde\nabla^\nu K^\mu{}_{\mu\lambda}+ \,\tilde\nabla_\nu \,\tilde\nabla_\mu K^{\nu\mu}{}_\lambda\right)+F(K,\, \,\tilde\nabla K;\, \,\tilde\nabla^2 \phi,\, \,\tilde\nabla \phi, \,\tilde R)\,,\label{eqn G4ddk}
\end{eqnarray}
where we have explicitly written down {\it all} of the second derivatives of contortion in ${\mathcal{E}}_{\phi} $, which come from the terms $G_4 \, \tilde\nabla K $ and $G_{4,X}\left(\, \tilde\nabla\phi\right) \left(\, \tilde\nabla \, \tilde\nabla\phi\right)K $ in the action (\ref{eqn G4Texplicitlag}),

\begin{dmath}
{\mathcal{E}}_{K^\mu{}_{\nu\sigma}} =G_{4,X}\, \left(\delta^\nu_\mu\left(\tilde{\nabla}^\sigma \tilde{\nabla}_\rho\phi \tilde{\nabla}^\rho\phi-\tilde{\nabla}_\rho\tilde{\nabla}^\rho\phi \tilde{\nabla}^\sigma\phi\right)+g^{\sigma\nu}\left(\tilde{\nabla}_\rho\tilde{\nabla}^\rho\phi \tilde{\nabla}_\mu\phi-\tilde{\nabla}_\mu \tilde{\nabla}_\rho\phi \tilde{\nabla}^\rho\phi\right)+ \tilde{\nabla}^\nu \tilde{\nabla}_\mu\phi \tilde{\nabla}^\sigma\phi-\tilde{\nabla}^\nu \tilde{\nabla}^\sigma\phi \tilde{\nabla}_\mu\phi \right)\\
+G_{4,\phi}\left(\tilde{\nabla}_\mu\phi\, g^{\sigma\nu}-\tilde{\nabla}^\sigma\phi \, \delta^\nu_\mu\right)+G_4\left(\delta^\nu_\mu K^\sigma\,_\rho\,^\rho-g^{\sigma\nu}\,K_{\mu\rho}\,^\rho+K_\mu\,^{\sigma\nu}+K^\nu\,_\mu\,^\sigma\right)\\
- (-1 + \, c) \,{G_{4,X}} \nabla^{\alpha }\phi \bigl(g_{\mu \alpha } (K^{\nu \sigma }{}_{\beta } -  K^{\sigma \nu }{}_{\beta }) \nabla^{\beta }\phi -  (K_{\alpha }{}^{\nu }{}_{\mu } + K^{\nu }{}_{\mu \alpha }) \nabla^{\sigma }\phi \bigr) \,.\label{eqn K}
\end{dmath}
where we stress on the dependance of ${\mathcal{E}}_{K^\mu{}_{\nu\sigma}} $ on second order terms $\tilde{\nabla}^2\phi$. And ${\mathcal{E}}_{g_{\mu\nu}} $ and $F$ are shown in the Appendix \ref{asec eqns}. 

All in all, there are three field equations (\ref{eqn fieldeoms}) with up to second order derivatives of {\it all} of the three fields. Therefore we could expect new degrees of freedom besides the usual tensor modes and the single scalar mode that are usually found in the Quartic Horndeski theory without torsion\footnote{In different formalisms such as Palatini it has also been pointed out that a non Christoffel connection can lead to new degrees of freedom in Hordeski theory \cite{kubota2021cosmological,arai2022cosmological}}.

The main objective of this work, in the second order formalism, is to precisely determine the  degrees of freedom for the theory (\ref{eqn G4Texplicitlag}). We explore this question at linearized order in a perturbative expansion.

\subsection{Modification of Horndeski degrees of freedom on a spacetime with torsion} \label{sec ssector}

Contrary to the expectation from the expression (\ref{eqn G4ddk}) there are in fact no explicit kinetic terms of contortion perturbations about a spatially flat FLRW background. This can be explicitly seen in the quadratic action for the tensor and scalar sectors, respectively (\ref{eqn ql0t}), (\ref{eqn ql0s}). On the other hand, the vector sector which is shown in the Appendix \ref{asec ql0}  is trivial in the sense that all vector perturbations are non-dynamical. The nontrivial part of the quadratic action is written as
\begin{equation}
\mathcal{S}_{4c}= \mathcal{S}^{Tensor} +\mathcal{S}^{Scalar}_c,
\end{equation}
where the part relevant to the three tensor perturbations $h_{ij},\, T^{\scalebox{0.5}{(1)}}_{ij},\, T^{\scalebox{0.5}{(2)}}_{ij} $ is

\begin{dmath}
\left.\mathcal{S}^{Tensor}=\right.  \frac{1}{2}\int\, \textrm{d}\eta\,\textrm{d}^3x \,\,\left( v_{1}\,(\dot{h}_{ij})^2+v_{2} \,(\partial_k h_{ij})^2+v_{3}\, (T^{\scalebox{0.5}{(1)}}_{ij})^2 + v_{4}\, (\partial_k T^{\scalebox{0.5}{(2)}}_{ij})^2 +v_5 \,h_{ij}\, T^{\scalebox{0.5}{(1)}}_{ij} + v_6  \,\dot{h}_{ij}\, T^{\scalebox{0.5}{(1)}}_{ij} + v_7 (h_{ij})^2 \right)\label{eqn ql0t}\,,  
\end{dmath}
where the coefficients $v_i$ with $i=1 \dots\, 7$, shown in the Appendix \ref{asec ql0}, depend only on background quantities. In particular, they are independent of the parameter of the theory $c$. Hence the tensor modes are the same within the familiy of Quartic Horndeki Cartan theories. On the other hand, let us also point out in advance that after integrating out the torsion perturbations $ T^{\scalebox{0.5}{(1)}}_{ij},\, T^{\scalebox{0.5}{(2)}}_{ij} $ and after using the equations of motion for the background fields the mass term for the graviton vanishes, similar as in the torsionless Horndeski theory.\bigskip

The part of the action relevant to the $13$ scalar perturbations $C^{\scalebox{0.5}{(n)}},\, \alpha,\, B,\, \psi,\, E,\, \Pi$ with $n=1, \dots , 8$, without fixing gauge is:

\begin{dmath}\mathcal{S}^{Scalar}_{c}\,\text{=}\,\frac{1}{2}\, \int\, \textrm{d}\eta\,\textrm{d}^3x\,  \Big(\,c\, \left( f_{7} \, \,{\partial_{i}B} \,{\partial_{i}C^{\scalebox{0.5}{(1)}}}\, + \, f_{51} \, \,({\partial_{i}B})^2+\, f_{52} \, \,({\partial_{i}C^{\scalebox{0.5}{(1)}}})^2\right)\,\nonumber\\+\left(\,f_{1} \, \,{{\alpha}} \,{{\Pi}} \, + \, f_{2} \, \,{{C^{\scalebox{0.5}{(3)}}}} \,{{\psi}} \, + \, f_{3} \, \,{{\alpha}} \,{{\psi}} \, + \, f_{4} \, \,{{\Pi}} \,{{\psi}} \, + \, f_{5} \, \,{{C^{\scalebox{0.5}{(3)}}}} \,{{\Pi}} \, + \, f_{6} \, \,{{C^{\scalebox{0.5}{(3)}}}} \,{{\alpha}} \, + \, f_{8} \, \,{\partial_{i}C^{\scalebox{0.5}{(3)}}} \,{\partial_{i}E} \, + \, f_{9} \, \,{\partial_{i}B} \,{\partial_{i}C^{\scalebox{0.5}{(3)}}} \, + \, f_{10} \, \,{\partial_{i}C^{\scalebox{0.5}{(2)}}} \,{\partial_{i}C^{\scalebox{0.5}{(3)}}} \, + \, f_{11} \, \,{\partial_{i}C^{\scalebox{0.5}{(4)}}} \,{\partial_{i}C^{\scalebox{0.5}{(5)}}} \, + \, f_{12} \, \,{\partial_{i}B} \,{\partial_{i}C^{\scalebox{0.5}{(6)}}} \, + \, f_{13} \, \,{\partial_{i}C^{\scalebox{0.5}{(1)}}} \,{\partial_{i}C^{\scalebox{0.5}{(6)}}} \, + \, f_{14} \, \,{\partial_{i}B} \,{\partial_{i}\Pi} \, + \, f_{15} \, \,{\partial_{i}E} \,{\partial_{i}\Pi} \, + \, f_{16} \, \,{\partial_{i}C^{\scalebox{0.5}{(1)}}} \,{\partial_{i}\Pi} \, + \, f_{17} \, \,{\partial_{i}C^{\scalebox{0.5}{(2)}}} \,{\partial_{i}\Pi} \, + \, f_{18} \, \,{\partial_{i}C^{\scalebox{0.5}{(3)}}} \,{\partial_{i}\Pi} \, + \, f_{19} \, \,{\partial_{i}C^{\scalebox{0.5}{(6)}}} \,{\partial_{i}\Pi} \, + \, f_{20} \, \,{\partial_{i}B} \,{\partial_{i}\alpha} \, + \, f_{21} \, \,{\partial_{i}E} \,{\partial_{i}\alpha} \, + \, f_{22} \, \,{\partial_{i}C^{\scalebox{0.5}{(2)}}} \,{\partial_{i}\alpha} \, + \, f_{23} \, \,{\partial_{i}C^{\scalebox{0.5}{(6)}}} \,{\partial_{i}\alpha} \, + \, f_{24} \, \,{\partial_{i}\alpha} \,{\partial_{i}\Pi} \, + \, f_{25} \, \,{\partial_{i}E} \,{\partial_{i}\psi} \, + \, f_{26} \, \,{\partial_{i}B} \,{\partial_{i}\psi} \, + \, f_{27} \, \,{\partial_{i}C^{\scalebox{0.5}{(2)}}} \,{\partial_{i}\psi} \, + \, f_{28} \, \,{\partial_{i}\Pi} \,{\partial_{i}\psi} \, + \, f_{29} \, \,{\partial_{i}\alpha} \,{\partial_{i}\psi} \, + \, f_{30} \, \,{\partial_{i}\partial_{j}C^{\scalebox{0.5}{(7)}}} \,{\partial_{i}\partial_{j}C^{\scalebox{0.5}{(8)}}} \, + \, f_{31} \, \,{{\psi}} \,{\dot{\Pi}} \, + \, f_{32} \, \,{{\alpha}} \,{\dot{\Pi}} \, + \, f_{33} \, \,{{C^{\scalebox{0.5}{(3)}}}} \,{\dot{\Pi}} \, + \, f_{34} \, \,{{\alpha}} \,{\dot{\psi}} \, + \, f_{35} \, \,{{C^{\scalebox{0.5}{(3)}}}} \,{\dot{\psi}} \, + \, f_{36} \, \,{\dot{\Pi}} \,{\dot{\psi}} \, + \, f_{37} \, \,{\partial_{i}C^{\scalebox{0.5}{(3)}}} \,{\partial_{i}\dot{E}} \, + \, f_{38} \, \,{\partial_{i}\alpha} \,{\partial_{i}\dot{E}} \, + \, f_{39} \, \,{\partial_{i}B} \,{\partial_{i}\dot{\Pi}} \, + \, f_{40} \, \,{\partial_{i}E} \,{\partial_{i}\dot{\Pi}} \, + \, f_{41} \, \,{\partial_{i}C^{\scalebox{0.5}{(2)}}} \,{\partial_{i}\dot{\Pi}} \, + \, f_{42} \, \,{\partial_{i}C^{\scalebox{0.5}{(6)}}} \,{\partial_{i}\dot{\Pi}} \, + \, f_{43} \, \,{\partial_{i}\dot{E}} \,{\partial_{i}\dot{\Pi}} \, + \, f_{44} \, \,{\partial_{i}B} \,{\partial_{i}\dot{\psi}} \, + \, f_{45} \, \,{\partial_{i}C^{\scalebox{0.5}{(2)}}} \,{\partial_{i}\dot{\psi}} \, + \, f_{46} \, \,{\partial_{i}\dot{E}} \,{\partial_{i}\dot{\psi}} \, + \, f_{47} \, \,({{C^{\scalebox{0.5}{(3)}}}})^2 \, + \, f_{48} \, \,{{\alpha}}^2 \, + \, f_{49} \, \,{{\psi}}^2 \, + \, f_{50} \, \,{{\Pi}}^2 \, \, + \, f_{53} \, \,({\partial_{i}C^{\scalebox{0.5}{(4)}}})^2 \, + \, f_{54} \, \,({\partial_{i}C^{\scalebox{0.5}{(6)}}})^2 \, + \, f_{55} \, \,({\partial_{i}\Pi})^2 \, + \, f_{56} \, \,({\partial_{i}\psi})^2 \, + \, f_{57} \, \,({\partial_{i}\partial_{j}C^{\scalebox{0.5}{(8)}}})^2 \, + \, f_{58} \, \,{\dot{\Pi}}^2 \, + \, f_{59} \, \,{\dot{\psi}}^2 \right)\Big) \, \,, \label{eqn ql0s} \end{dmath}

where the coefficients $f_i$ with $i=1 \dots\, 59$ depend only on background quantities and the parameter of the theory $c$. They are presented in the Appendix \ref{asec ql0}.

With respect to the difference in dynamics within the family of Quartic Horndeski Cartan Lagrangians, it is important to notice that in the scalar sector the theory with $c=0$ has three less terms, as shown in the first line in (\ref{eqn ql0s}). In particular, {\it only} for the $c=0$ theory, the metric perturbation $B$ and the first torsion scalar $C^{\scalebox{0.5}{(1)}} $ are Lagrange multipliers. As we show below, these constraints are critical in the sense that they lead to different dynamics of the scalar field perturbation in the $c=0$ theory compared to the Quartic Horndeski Cartan theories with non zero parameter $c$.\bigskip

Even though there are no explicit kinetic terms for the contortion tensor and scalar perturbations in (\ref{eqn ql0t}) and (\ref{eqn ql0s}), there could still arise new kinetic terms after using some of the equations of motion. As we show below this is not the case for the Quartic Horndeski Cartan theories (\ref{eqn G4Texplicitlag}). Namely, for the scalar and tensor sectors on the spatially flat FLRW background (\ref{eqn ql0t}), (\ref{eqn ql0s}) it is possible to use all constraint equations to integrate out {\it all} torsion perturbations $C^{\scalebox{0.5}{(n)}} $ ($n=1,\dots 8$), $T^{\scalebox{0.5}{(1)}}_{ij}, \, T^{\scalebox{0.5}{(2)}}_{ij} $ as well as the metric perturbations $\alpha,\, B$ before fixing the gauge. To obtain a final result we assume  that the  background scalar field $\varphi(\eta)$ is not constant ($\dot{\varphi}\neq 0$), that the torsion background $x$ is non vanishing and that coefficients in the action (\ref{eqn ql0t}) and (\ref{eqn ql0s}) and after using constraint equations do not vanish\footnote{This clearly restricts even the most general result to a class of $G_4$ functions with some non vanishing derivatives. For instance, we assume that $G_{4,\phi}\neq 0$ and $G_{4,X}\neq 0$. We show a concrete example in section \ref{sec example}. We also assume a non trivial branch of solutions for the background fields (\ref{eqn backgroundeoms}). For instance, we assume that $x+ a\,\dot{a}\neq 0$, which is a very particular branch of solutions to the background equation $\mathcal{E}_{g_{00}}=0$ (\ref{eqn backgroundeoms}) that is  trivial in the sense that it vanishes many coefficients in the quadratic Lagrangian.}. All in all, the second order action in the unitary gauge, where $\Pi=0$ and $E=0$, divides only into tensor modes $h_{ij}$ and a single scalar mode $\psi$, thus showing that generally, a torsionful connection on the spacetime does not introduce additional degrees of freedom but it only modifies the usual tensor and scalar degrees of freedom that are  found in the standard, torsionless Quartic Horndeski Lagrangian. The final form of the action for general $c$ parameter reads
\begin{eqnarray}
S_{4c}=\frac{1}{2}\int\, \textrm{d}\eta\,\textrm{d}^3x \,a^4\,\left[\frac{1}{\,a^2}\left({\mathcal{G}_\tau}\left(\dot{h}_{ij}\right)^2-{\mathcal{F}_\tau}(\partial_k\,{h}_{ij})^2\right) + \frac{1}{a^2}\left(\dot{\psi}\left(\mathcal{G}_{\mathcal{S}\textrm{I}}- c\,\frac{1}{a^2}\, \mathcal{G}_{\mathcal{S}\textrm{II}}\,\partial_i \partial_i \right)\dot{\psi}-\mathcal{F}_{\mathcal{S}}(\partial_i \psi)^2 \right)\right],\label{eqn finalql}
\end{eqnarray}
where
\begin{eqnarray} 
\mathcal{G}_\tau &=& 2\frac{\, G_4^2}{\,G_4+\,2\,X \, G_{4,X}}\,,\\
\mathcal{F}_\tau &=&2\,G_4\,,\\
\mathcal{G}_{\mathcal{S}\textrm{I}}&=& \frac{m_1}{a^2\,(- \,{G_{4,{\phi}}} \,{G_{4,X}} + \,{G_{4,{\phi}X}} \,{G_{4}}) \bigl(\,{a}^4 \,{G_{4}}^2 + (2 \,{G_{4,X}}^2 -  \,{G_{4,XX}} \,{G_{4}}) \,\dot{\varphi}^4\bigr)^2}\,,\\
\mathcal{G}_{\mathcal{S}\textrm{II}}&=&\frac{8 \,{G_{4,X}}^3 \,{G_{4}}^3}{\left(\,G_{4}+c\,X\,G_{4,X}\right)\left(G_{4,X}\, G_{4,\phi}-G_{4}\, G_{4,\phi\,X}\right)^2}\,,\label{eqn gs2general}\\
\mathcal{F}_{\mathcal{S}}&=& \frac{m_2}{m_3}\,,
\end{eqnarray}
where the background functions $m_i$ with $i=1, \,2,\, 3$ are not very illuminating and are shown in the Appendix \ref{sec appfinalql}.\bigskip

The crucial aspect evident in the action (\ref{eqn finalql}) is the parameter of the theory $c$ in the coefficient of the term $\dot{\psi}\partial_i \partial_i \dot\psi$. This contribution radically modifies the dispersion relation of the scalar in the case it does not vanish. Hence, this result singles out the Quartic Horndeski Cartan theory (\ref{eqn G4Tlag}) with $c=0$ as the only one with a scalar degree of freedom with a regular wave-like behavior. In such a case the scalar sector is {\it not necessarily} the same in the torsionless and the torsionful Quartic Horndeski theory, because the speed of sound squared
\begin{eqnarray}
\left(c^{HC}_{s}\right)^2= \frac{\mathcal{F}_{\mathcal{S}}}{\mathcal{G}_{\mathcal{S}\textrm{I}}}\,,
\end{eqnarray}
is different in both cases (See for instance section $3.2.1$ in \cite{kobayashi2019horndeski}).\bigskip

For the theories $\mathcal{S}_{4c}$ with $c\neq 0$ there are some crucial aspects different in comparison to the theory with $c=0$. Namely, when the parameter of the theory $c$ is non vanishing: $1)\,$ $ C^{\scalebox{0.5}{(1)}} $ is {\it not} a Lagrange multiplier. $2)\,$ $C^{\scalebox{0.5}{(1)}} $ and $B$ are coupled. $3)\,$ The scalar $B$ is {\it not} a Lagrange multiplier, as opposed to the standard Quartic Horndeski theory on a torsionless spacetime. All these three aspects are evident from the first line in expression (\ref{eqn ql0s}). Let us clarify how these features finally amount to the disparity in the propagation of the scalar mode in theories with $c=0$ and $c\neq 0$: the essential aspects that we discuss in the scalar sector in expression (\ref{eqn finalql}) can be written in a toy model with a Lagrangian of the form
\begin{eqnarray}
\mathcal{L}_{toy}=\left(1-c\,p^2\right)\dot{x}_1^2-p^2\,x_1^2\,,  \label{eqn toy2}
\end{eqnarray}
for a field $x_1$ in momentum space. This toy model has the property that if $c\neq 0$ the field $x_1$ has an unusual dispersion relation of the form discussed before. However, there is an equivalent toy model to (\ref{eqn toy2})
\begin{eqnarray}
\mathcal{L}_{toy'}=\dot{x}_1^2-p^2\,x_1^2+ c\,p^2\, \left( 2\,x_2\,\dot{x}_1+ x_2^2\right)\,,\label{eqn toy}
\end{eqnarray}
which {\it off-shell} seems to be the Lagrangian for a degree of freedom $x_1$ with a standard wave-like dispersion relation, but coupled to an additional auxiliary field $x_2$. Indeed, plugging back in the Lagrangian (\ref{eqn toy}) the equation of motion for $x_2$ $(x_2=-\dot{x}_1)$ we recover the toy model (\ref{eqn toy2}).

For the theories $\mathcal{S}_{4c}$ there is a simplified analogy with the toy model (\ref{eqn toy}),  which follows identifying $\psi$ and $C^{\scalebox{0.5}{(6)}} $ with $x_1$ and $x_2$, respectively. Indeed, on one hand, the terms of the toy model (\ref{eqn toy}), namely $\dot{\psi}^2,\,p^2\psi^2,\,p^2(C^{\scalebox{0.5}{(6)}})^2$ are present in the initial action (\ref{eqn ql0s}). On the other hand, the remaining key term to complete the analogy to the toy model (\ref{eqn toy}),  namely $p^2 C^{\scalebox{0.5}{(6)}} \dot{\psi}$, is generated by means of the term $\, p^2 \,f_{23}\, C^{\scalebox{0.5}{(6)}} \alpha $ in the initial action (\ref{eqn ql0s}) because on-shell (up to background dependent coefficients)
\begin{eqnarray}
\alpha=\dot{\psi}+\dots\,.
\end{eqnarray}
Now, it is clear that in order to obtain the unusual dispersion relation in the toy model (\ref{eqn toy}) $x_2$ must not vanish. Analogously, the auxiliary field $C^{\scalebox{0.5}{(6)}} $ must not vanish. This is only the case when $c\neq 0$. Indeed, the equation of motion for the first torsion scalar $C^{\scalebox{0.5}{(1)}} $ computed from the action (\ref{eqn ql0s}) in the unitary gauge for a theory with general parameter $c$ is
\begin{eqnarray}
\, C^{\scalebox{0.5}{(6)}}=-c\,\frac{1}{f_{13}}\,\left(2\,f_{52}\,C^{\scalebox{0.5}{(1)}}\,+\,f_{7}\,B\right)\,.\label{eqn c1}
\end{eqnarray}
{\it Only } if $c\neq 0$ the first torsion scalar $C^{\scalebox{0.5}{(1)}} $ is not a Lagrange multiplier (let us here recall the first critical difference stated above between theories with zero and non zero $c$ parameter) and therefore $C^{\scalebox{0.5}{(6)}} $ does not vanish. As a consequence, in analogy with the equivalence between toy models (\ref{eqn toy}) and (\ref{eqn toy2}), when $c\neq 0$ we obtain a non regular wave-like dispersion relation as we found in the result (\ref{eqn finalql}). 

This scalar degree of freedom for theories with $c\neq 0$ has no counterpart with the Quartic Horndeski theory on a torsionless spacetime. We explore further this unusual dispersion relation for the theories with $c\neq 0$ with a concrete example in section \ref{sec example}. \bigskip

Finally, for the tensor modes the sound speed squared is modified from the standard (torsionless) Quartic Horndeski case, from \cite{kobayashi2019horndeski}
\begin{eqnarray}
\left(c^{H}_{\tau}\right)^2=\frac{G_4}{G_4-2\,X\,G_{4,X}}\,,
\end{eqnarray}
to
\begin{eqnarray}
\left(c^{HC}_{\tau}\right)^2=\frac{ \mathcal{F}_\tau}{\mathcal{G}_\tau}=\left(1+2\,X\frac{G_{4,X}}{G_4}\right)\,.\label{eqn soundspeedtensor}
\end{eqnarray}
The difference between $c^{H}_{\tau} $ and $c^{HC}_{\tau} $ boils down to the torsion perturbation $T^{\scalebox{0.5}{(1)}}_{ij}$, which couples to $h_{ij}$. On the other hand, $T^{\scalebox{0.5}{(2)}}_{ij}$ does not couple to the graviton. Let us note that this result on the tensor sector is independent of the torsion background $x(\eta)$.

\subsection{Stability and classification of the scalar mode in Horndeski Cartan theories}\label{sec classification}

The stability conditions to avoid ghost and gradient instabilities for the tensor modes are
\begin{equation}
\begin{array}{cc}
\mathcal{G}_{\tau}>0\,,& \mathcal{F}_{\tau}>0\,, \label{eqn stabilitytensor}
\end{array}
\end{equation}
which are independent of the parameter of the theory $c$. Expressions (\ref{eqn stabilitytensor}) can be written altogether with the requirement of subluminality of the graviton $\left(c^{HC}_{\tau}\right)^2 <1$ as
\begin{equation}
\begin{array}{c}
 G_4>-2\,X\,G_{4,X}>0\,, \label{eqn gravitonstability}
\end{array}
\end{equation}
and easily compared the analogous conditions for the tensor modes in the torsionless Quartic Horndeski theory \cite{kobayashi2019horndeski}
\begin{equation}
\begin{array}{c}
G_4>0>2\,X\,G_{4,X}\,.
\end{array}
\end{equation}
 
On the other hand, for the scalar mode we require a separate analysis for the theories with $c=0$ and non zero $c$.\bigskip

\paragraph{Stability in the scalar mode for the theory $\mathcal{S}_{4c}$ with $c=0$}

In this theory the scalar mode has the usual wave-like dispersion relation and the ghost-free and stability conditions are as usual
\begin{equation}
\begin{array}{cc}
\mathcal{G}_{\mathcal{S} I}>0\,,& \mathcal{F}_{\mathcal{S}}>0\,,
\end{array}
\end{equation}
provided that  we have also required a positive sign for the kinetic term of the graviton ($\mathcal{G}_{\tau}>0 $).\bigskip

\paragraph{Ghost-free condition in the scalar mode for the theories $\mathcal{S}_{4c}$ with $c\neq 0$} In this case, the non wave-like dispersion relation only allows to state the ghost-free condition as
\begin{equation}
\begin{array}{cc}
\mathcal{G}_{\mathcal{S} I}>0\,,& c\,\mathcal{G}_{\mathcal{S} II}>0\,.
\end{array}
\end{equation}
Also, drawing an analogy with an oscillatory system with a restaurative force and bounded energy from below, we can demand a similar condition to gradient stability
\begin{equation}
\begin{array}{c}
\mathcal{F}_{\mathcal{S}}>0\,.
\end{array}
\end{equation}
To advance further in the analysis let us consider the case of high momentum. Then, only $\mathcal{G}_{\mathcal{S} II} $ and $\mathcal{F}_{\mathcal{S}} $ are relevant. In such a case, from equation (\ref{eqn gs2general}) the no ghost  condition for the scalar reduces to 
\begin{eqnarray}
c\,\mathcal{G}_{\mathcal{S} II}=\,\frac{8\, c \,{G_{4,X}}^3 \,{G_{4}}^3}{\left(\,G_{4}+c\,X\,G_{4,X}\right)\left(G_{4,X}\, G_{4,\phi}-G_{4}\, G_{4,\phi\,X}\right)^2}&>&0\,. \label{eqn nonghosthighpscalar}
\end{eqnarray}
Assuming the stability, subluminality and no ghost condition for the graviton (\ref{eqn gravitonstability}), which explicitly restricts to 
\begin{equation}\begin{array}{cc}G_4>0\,, & G_{4,X}<0\,,\end{array}\end{equation} 
because for the background fields $X=\frac{1}{2\,a^2}\dot{\varphi}^2\,>\,0$, we can rewrite the no ghost condition for the scalar (\ref{eqn nonghosthighpscalar}), as
\begin{eqnarray}
\frac{c}{\left(\,G_{4}+c\,X\,G_{4,X}\right)}&<&0\,.\label{eqn nonghosthighpscalar2}
\end{eqnarray}
Using again the stability of the graviton (\ref{eqn gravitonstability}) rewritten as
\begin{eqnarray}
\left(\,G_{4}+c\,X\,G_{4,X}\right)&>&(c-2)\,X\,G_{4,X}
\end{eqnarray}
we find from (\ref{eqn nonghosthighpscalar2}) that the scalar degree of freedom for theories $\mathcal{S}_{4c}$ with $c\neq 0$ can be classified in the {\it high momentum approximation } as non ghost for theories with $c<0$, and as a ghost if $0<c\leq 2$. Hence, we reach the important conclusion that there are Horndeski Cartan theories ($c<0$) for which both the graviton and the scalar mode are {\it simultaneously} ghost-free in the high momentum limit.

Let us stress that this classification of the scalar mode only holds by simultaneously assuming that  the graviton is subluminal, stable and ghost-free (\ref{eqn gravitonstability}) and cannot be deduced by stability considerations of the scalar mode on its own. Table \ref{table classification} summarizes these results.

The stability of the scalar mode in a particular example with $c\neq 0$ is further examined in section \ref{sec example}.

\section{Example: The scalar mode in Horndeski Cartan theories $\mathcal{S}_{4c}$ with nonzero $c$}\label{sec example}

\begin{figure}[b]
 \centering
\begin{subfigure}{0.55\textwidth}
\centering
  \includegraphics[width=1\linewidth]{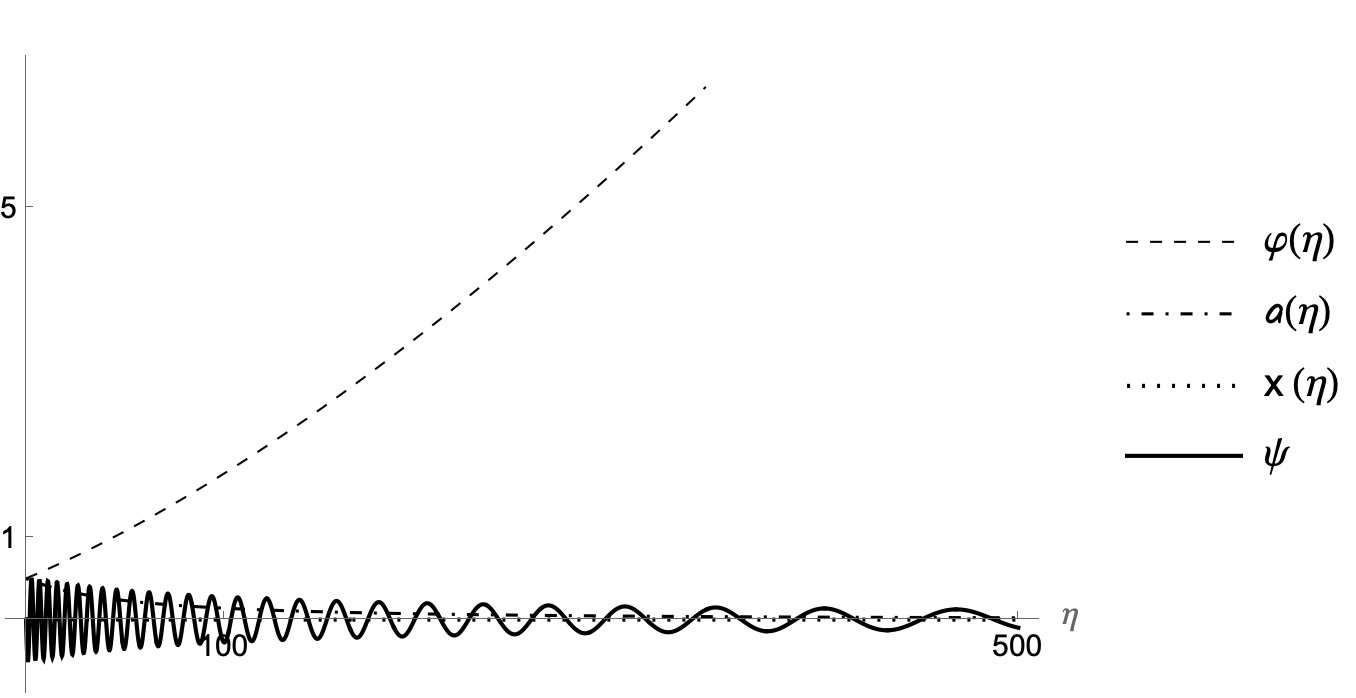}
 \caption{$\vert\vec{p}\vert\rightarrow\infty$} \label{fig ic stable}
\end{subfigure}%
\begin{subfigure}{0.45\textwidth}
\centering
  \includegraphics[width=1\linewidth]{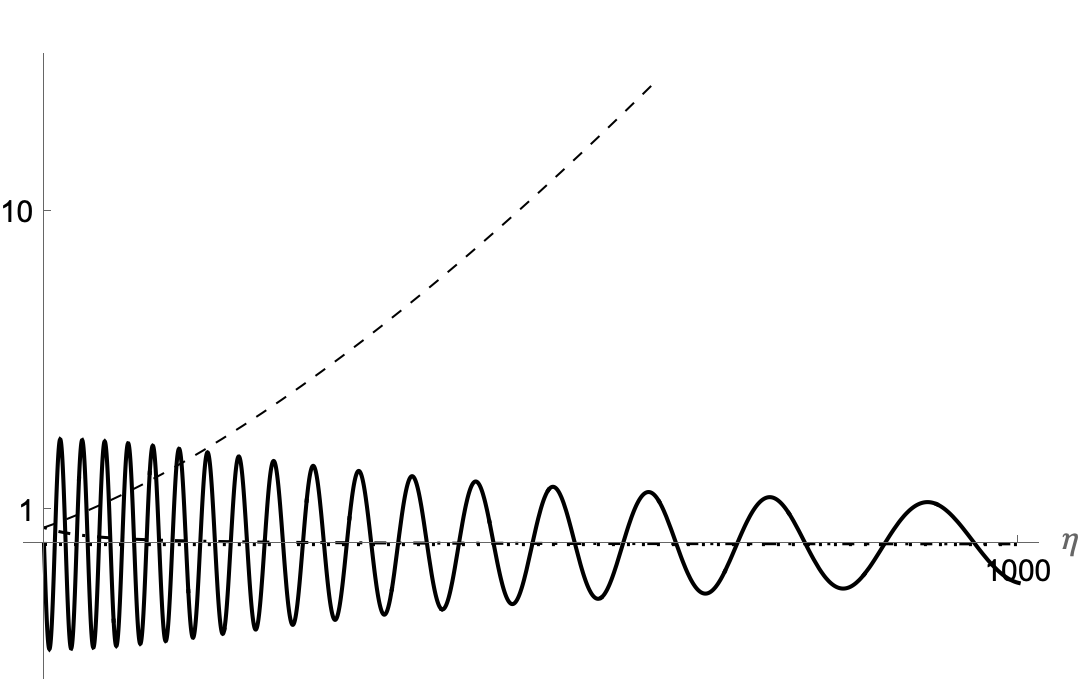}
 \caption{$\vert\vec{p}\vert=0.5$} \label{fig ic approxstable}
\end{subfigure}\caption{Evolution of the scalar mode $\psi$ for widely different momenta. In both cases the amplitude of oscillation of the scalar mode is decreasing, approaching zero before a late time when the assumptions are violated and a singularity is developed at $\eta\approx 3983$. The position of this singularity (not shown in the figures) is associated to the time when $a\rightarrow 0$ and independent of the momentum. The initial data is such that the conditions for a healthy graviton (\ref{eqn stgravex}) are met at the initial time and also they are such that $\varphi\dot\varphi\vert_{\eta=0}>0 $, such that the assumptions $\varphi(\eta)\neq 0$,\, $\dot{\varphi}(\eta)\neq 0$ are satisfied for the time domains of numerical evolution (Notice that $\varphi(\eta)$ seems monotonic increasing within these time domains). Initial Conditions $\varphi(0)=0.5,\,\dot{\varphi}(0)=0.01,\, a(0)=0.5,\, \psi(0,|\vec{p}|)=0,\, \dot{\psi}(0,|\vec{p}|)=-0.9 $.}
\end{figure}

We have shown that only the Quartic Horndeski Cartan theory $\mathcal{S}_{4\, c}$ with $c=0$ has a scalar mode that propagates with a regular wave-like dispersion relation on the FLRW background. For other theories, namely, when $c\neq 0$, we showed that the dispersion relation is unusual. In this section we consider an example for a theory with $c\neq 0$ and we observe that the unusual dispersion relation does not necessarily imply an instability. 

Let us consider a $G_4$ function that can satisfy the subluminality, ghost-free and stability conditions of the graviton (\ref{eqn gravitonstability})
\begin{eqnarray}
G_4=\phi^2-2 \,\kappa\,X\,,\label{eqn g4ex}
\end{eqnarray}
where, in natural units $\kappa$  is a constant with dimension of length squared. For simplicity we work in Planck units, so, we choose $\kappa=1$.  
Furthermore, let us consider a theory with negative $c$ parameter, $c=-\frac{1}{2}$, which, as discussed in the last section, has the property that whenever the healthy graviton conditions (\ref{eqn gravitonstability}) are met, then the scalar is not a ghost in the high momentum approximation (See also Table \ref{table classification}).
    
Using the final form of the action (\ref{eqn finalql}) with the $G_4$ choice (\ref{eqn g4ex}) and for the theory $c=-\frac{1}{2}$, the action of the scalar sector is
\begin{eqnarray}
S_{4\,c}^{Scalar}=\frac{1}{2}\int\,\text{d}\eta\,\text{d}^3x\,a^2\left(\mathcal{G}_{\mathcal{S}\textrm{I}}\, \dot{\psi}^2+ c\,\frac{1}{a^2}\, \mathcal{G}_{\mathcal{S}\textrm{II}}\,(\partial_i \dot{\psi})^2-\mathcal{F}_{\mathcal{S}}(\partial_i \psi)^2 \right)\,,\label{eqn finalqlex}
\end{eqnarray}
where
\begin{eqnarray}
\mathcal{G}_{\mathcal{S}\textrm{I}}=\frac{24 (\,{a}^2 \,{\varphi}^2 - 3 \,\dot{\varphi}^2) (- \,{a}^2 \,{\varphi}^2 + \,\dot{\varphi}^2)^2 (3 \,{a}^4 \,{\varphi}^4 + 2 \,{a}^2 \,{\varphi}^2 \,\dot{\varphi}^2 + 3 \,\dot{\varphi}^4)}{a^2\,(\,{a}^4 \,{\varphi}^4 - 2 \,{a}^2 \,{\varphi}^2 \,\dot{\varphi}^2 + 9 \,\dot{\varphi}^4)^2}\,,
\end{eqnarray}
\begin{eqnarray}
c\,\mathcal{G}_{\mathcal{S}\textrm{II}}= \frac{4 (\,{a}^2 \,{\varphi}^2 -  \,\dot{\varphi}^2)^3}{a^4\,\varphi^2\,(2\,a^2\,\varphi^2-\dot{\varphi}^2)}\,,
\end{eqnarray}
\begin{eqnarray}
\mathcal{F}_{\mathcal{S}}=  \frac{8 (\,{a}^2 \,{\varphi}^2 -  \,\dot{\varphi}^2) \bigl(27 \,\dot{\varphi}^8 + \,{a}^2 \,{\varphi}^2 \,\dot{\varphi}^4 (28 \,{a}^2 \,{\varphi}^2 - 15 \,\dot{\varphi}^2) + \,{a}^6 \,{\varphi}^6 (3 \,{a}^2 \,{\varphi}^2 - 11 \,\dot{\varphi}^2)\bigr)}{\,{a}^2 \,{\varphi}^2 (\,{a}^2 \,{\varphi}^2 - 3 \,\dot{\varphi}^2) \bigl(9 \,\dot{\varphi}^4 + \,{a}^2 \,{\varphi}^2 (\,{a}^2 \,{\varphi}^2 - 2 \,\dot{\varphi}^2)\bigr)}\,.
\end{eqnarray}
Let us notice that the scalar is not a ghost whenever the graviton is healthy: namely, the subluminality, stability and ghost-free condition for the graviton (\ref{eqn gravitonstability}) is satisfied for time domains where 
\begin{eqnarray}
a^2\, \varphi^2-3\,\dot{\varphi}^2>0\,.\label{eqn stgravex}
\end{eqnarray}
When the graviton is {\it healthy} the scalar is not a ghost because the inequality (\ref{eqn stgravex}) implies 

\begin{eqnarray}
\mathcal{G}_{\mathcal{S}\textrm{I}}&>&0 \,, \label{eqn nog1}\\
c\,\mathcal{G}_{\mathcal{S}\textrm{II}}&>&0\,.\label{eqn nog2}
\end{eqnarray} 

Furthermore, in accordance with our assumptions while deriving (\ref{eqn finalql}), the torsion background does not vanish where the graviton is healthy (\ref{eqn stgravex}) and where $\phi,\,a$ and $\dot{\varphi} $ do not vanish, which were also part of the assumptions:
\begin{eqnarray}
 x = \frac{\,{a}^4 \,{\varphi} \,\dot{\varphi} \left(\,{a}^2 \,{\varphi}^2 -  \,\dot{\varphi}^2\right)}{9 \,\dot{\varphi}^4 + \,{a}^2 \,{\varphi}^2 (\,{a}^2 \,{\varphi}^2 - 2 \,\dot{\varphi}^2)} \,.\label{eqn x example}
\end{eqnarray}

Now, from (\ref{eqn finalqlex}) we can compute the equation of motion for $\psi(\eta,\vec{x})$. The coupled system of equations consists of one partial differential equation for $\psi$, with coefficients that depend only on time, and two ordinary differential equations for $\varphi,\,a$
\begin{eqnarray}
\ddot{\varphi}(\eta)-\dot{\varphi}^2\, \frac{\,{a}^8 \,{\varphi}^8 (\,{a}^2 \,{\varphi}^2 - 11 \,\dot{\varphi}^2) - 27 \,\dot{\varphi}^8 (5 \,{a}^2 \,{\varphi}^2 - 3 \,\dot{\varphi}^2) + 2 \,{a}^4 \,{\varphi}^4 \,\dot{\varphi}^4 (23 \,{a}^2 \,{\varphi}^2 + 25 \,\dot{\varphi}^2)}{2 \,{a}^4 \,{\varphi}^5 \,\dot{\varphi}^4 (19 \,{a}^2 \,{\varphi}^2 - 33 \,\dot{\varphi}^2) + \,{a}^8 \,{\varphi}^9 (3 \,{a}^2 \,{\varphi}^2 - 13 \,\dot{\varphi}^2) - 9 \,{\varphi} \,\dot{\varphi}^8 (\,{a}^2 \,{\varphi}^2 + 9 \,\dot{\varphi}^2)}&=&0\,, \label{eqn phiex}
\end{eqnarray}
\begin{eqnarray}
\,\dot{a} + \frac{\,{a}^3 \,{\varphi} \,\dot{\varphi} (\,{a}^2 \,{\varphi}^2 + 3 \,\dot{\varphi}^2)}{9 \,\dot{\varphi}^4 + \,{a}^2 \,{\varphi}^2 (\,{a}^2 \,{\varphi}^2 - 2 \,\dot{\varphi}^2)}&=&0\,, \label{eqn aex}
\end{eqnarray}

To solve the system numerically we consider the cases in high and low momentum. We also restrict the evolution to time domains where the graviton is healthy (\ref{eqn stgravex}), hence when (\ref{eqn nog1}) and (\ref{eqn nog2}) also hold and the scalar is not a ghost, and where the gradient stability condition $\mathcal{F}_{\mathcal{S}}>0$ holds. 

The numerical solutions in Figures \ref{fig ic stable}, \ref{fig ic approxstable}  suggest that the amplitude of oscillation of the scalar perturbation $\psi$  is decreasing and oscillates close to zero for late times, for a wide range of momenta. This behavior was observed restricted to initial data that satisfies the conditions for a healthy graviton (\ref{eqn stgravex}) and for as long as the assumptions hold. In particular there is a singularity not shown in the Figures \ref{fig ic stable}, \ref{fig ic approxstable} that only develops at late times, which is associated with the vanishing of $a$.

\begin{figure}[t]
 \centering
  \includegraphics[width=0.8\textwidth]{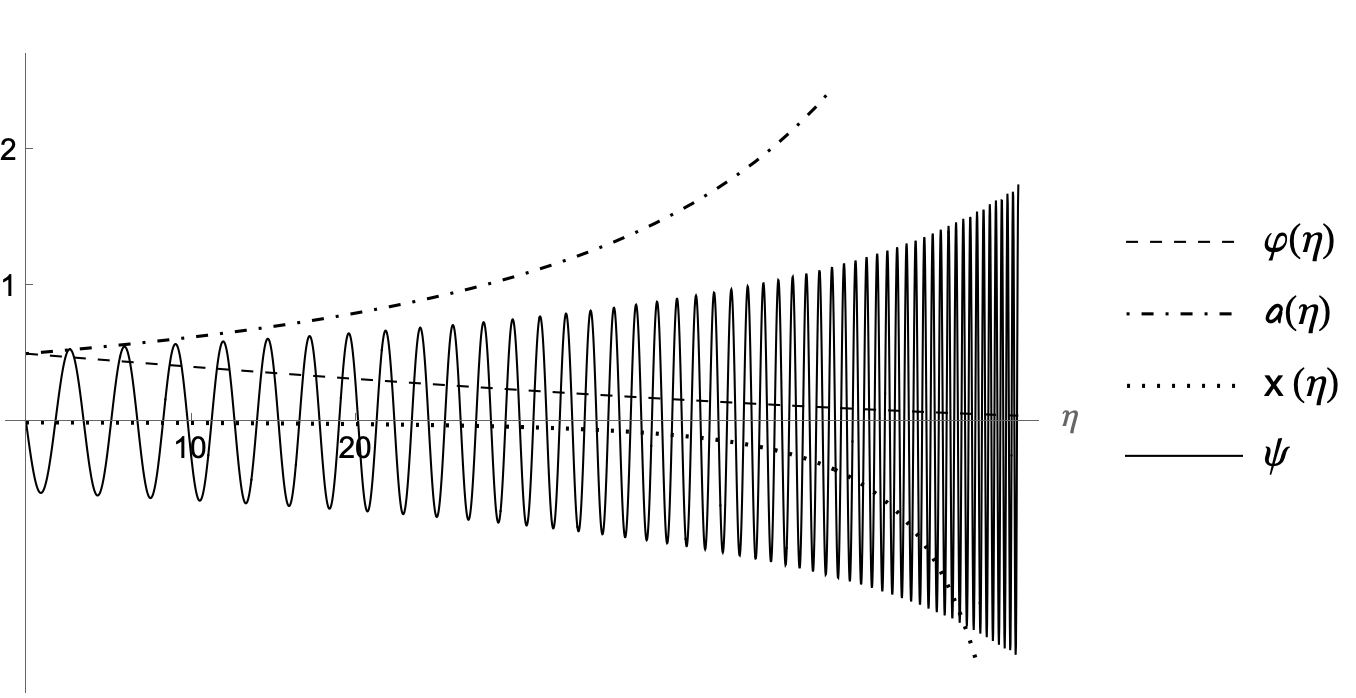}
 \caption{Typical case where a singularity quickly develops at the point where the scalar background $\varphi$ vanishes. Such scenarios are commonly obtained when the initial value of $\varphi$ and its velocity have opposite signs ($\varphi\dot{\varphi}\vert_{\eta=0}<0 $), because $\varphi$ is monotonic in the tested time domains. The vanishing of $\varphi$ violates the assumptions for the derivation of the action (\ref{eqn finalqlex}) (High momentum limit. Initial Conditions $\varphi(0)=0.5,\,\dot{\varphi}(0)=-0.01,\, a(0)=0.5,\, \psi(0,|\vec{p}|\rightarrow \infty)=0,\, \dot{\psi}(0,|\vec{p}|\rightarrow \infty)=-0.9 $. Singularity at $\eta\approx 75$).} \label{fig unstable}
\end{figure}

On the other hand, Figure \ref{fig unstable} shows a typical case where a singularity  quickly develops at early times\footnote{Early in comparison to the case in Figure \ref{fig ic stable}. Namely, $\eta_{\text{singularity}}\sim 10$ in Figure \ref{fig unstable} compared to $\eta_{\text{singularity}}\sim 10^3$ in Figure \ref{fig ic stable}. }, because the initial conditions quickly lead to violate the assumption $\dot{\varphi}\neq 0$ upon which the validity of the result (\ref{eqn finalqlex}) and figures rely. Approaching this singularity seems to lead to a growing amplitude of the scalar perturbation, even though the gradient stability condition $\mathcal{F}_{\mathcal{S}}>0$ holds in the tested time domain in  Figure \ref{fig unstable}.  Thus, we avoid initial conditions of the scalar that quickly lead to $\dot{\varphi}= 0$ and singularities. More precisely, because $\varphi$ is monotonic {\it at least within the tested  time domain}, we can easily avoid this singularity by restricting to initial conditions of the scalar that satisfy $\varphi\dot\varphi\vert_{\eta=0}>0 $.

All in all, despite the unusual dispersion relation for the Quartic Horndeski Cartan theories with $c\neq0$, the results in this section for $c=-\frac{1}{2}$ suggest that the scalar mode is not necessarily unstable because at least in some cases and within the tested time domains, which we chose to coincide with domains where the graviton is also healthy, the amplitude of oscillation of the scalar perturbation decreases with time, settling to zero for late times, whenever the initial conditions do not lead to violate the assumptions upon which the computations were derived.
\section{Conclusion}\label{sec conclusions}

We considered the Quartic Horndeski theory with torsion in the second order formalism, written as a one parameter $(c)$ family 
\begin{eqnarray}
\mathcal{S}_{4c}= \int \,\text{d}^4x\,\sqrt{-g}\,\left(\, G_4(\phi,X)\tilde{R}+G_{4,X}\left(\left(\tilde{\nabla}_\mu\tilde{\nabla}^\mu\phi\right)^2-\left(\tilde{\nabla}_\mu\tilde{\nabla}_\nu\phi\right) \tilde{\nabla}^\nu\tilde{\nabla}^\mu\phi -c\,\left(\tilde{\nabla}_\mu\tilde{\nabla}_\nu\phi\right) \left[\tilde{\nabla}^\mu,\tilde{\nabla}^\nu\right]\phi \right)\right)\, ,\label{eqn G4Tlag concl}
\end{eqnarray} 
where tilde denotes torsionful quantities and the theories (\ref{eqn G4Tlag concl}) for all $c$ reduce to the standard Horndeski action without torsion by taking $\tilde{R}\rightarrow R,$ and $ \tilde{\nabla}\rightarrow \nabla$ such that $\left[{\nabla}^\mu,{\nabla}^\nu\right]\phi=0 $, where $\nabla$ denotes covariant derivative defined with a Christoffel connection. 

We showed in a perturbative expansion that these Quartic Horndeski Cartan theories at linear order on a spatially flat FLRW background do not introduce additional degrees of freedom and that the torsionful connection only modifies the usual tensor and scalar degrees of freedom when compared to the standard Quartic Horndeski theory without torsion. Indeed, the quadratic action for the one parameter family of theories (\ref{eqn G4Tlag concl}) can be finally written for the graviton $h_{ij}$ and a single scalar field perturbation $\psi$ in the form
\begin{eqnarray}
S_{4c}=\frac{1}{2}\int\, \textrm{d}\eta\,\textrm{d}^3x \,a^4\,\left[\frac{1}{\,a^2}\left({\mathcal{G}_\tau}\left(\dot{h}_{ij}\right)^2-{\mathcal{F}_\tau}(\partial_k\,{h}_{ij})^2\right) + \frac{1}{a^2}\left(\dot{\psi}\left(\mathcal{G}_{\mathcal{S}\textrm{I}}- c\,\frac{1}{a^2}\, \mathcal{G}_{\mathcal{S}\textrm{II}}\,\partial_i \partial_i \right)\dot{\psi}-\mathcal{F}_{\mathcal{S}}(\partial_i \psi)^2 \right)\right]\,,\label{eqn finalql concl}
\end{eqnarray}
where $\mathcal{G}_\tau,\, \mathcal{F}_\tau,\, \mathcal{G}_{\mathcal{S}\textrm{I}} ,\, \mathcal{G}_{\mathcal{S}\textrm{II}} ,\, \mathcal{F}_{\mathcal{S}} $ are functions of $G_4$ and its derivatives.

We computed the speed of sound for the graviton, which is the same in all of the Quartic Horndeski Cartan theories, and found that the subluminality, ghost-free and stability conditions are similar to the standard Horndeski theory without torsion, as shown in Table \ref{table classification}. We also showed that for the theories (\ref{eqn G4Tlag concl}) with parameter $c\neq 0$ the dispersion relation of the scalar mode is radically modified and it has no counterpart with the usual scalar mode in the torsionless Horndeski theory, as can be seen in the term proportial to $c$ in expression (\ref{eqn finalql concl}). We analyzed the latter in a particular example and observed that the unusual non wave-like dispersion relation when $c\neq 0$ does not necessarily imply an instability. Furthermore, we showed that there are Horndeski Cartan theories ($c < 0$) for which both the graviton and the scalar mode are simultaneously ghost-free in the high momentum limit, conditional to the assumption that the graviton is also stable and subluminal.  

We found that the theory with parameter $c=0$ is the only one within the family in which the scalar field perturbation propagates with a regular wave-like dispersion relation and yet,  it is {\it not necessarily} the same in comparison to the torsionless Quartic Horndeski theory because its speed of sound is different.

\begin{table*}[h]
\centering
    \large        
    \caption{Summary of tensor and scalar modes classified according to the parameter $c$ of the theory. ${}^{*}$We refer to a stable, non ghost and subluminal graviton as {\it healthy}.  The ghost free conditions in the scalar mode are only written if they are tied to the assumption of a healthy graviton. Otherwise they require further assumptions depending on $G_4$.  }\label{table classification}
    \centering    
    \begin{tabular}{c|c|c|c|c}
    \hline
     {}&$c<0$&$c=0$& $0<c\leq 2$&$c>2$\\ 
\hline
Scalar mode& \makecell{Non wave-like\\ dispersion relation.\\ {\it Not a ghost}\\ (in high momentum)  \\ if the graviton\\ is healthy${}^{*}$.}&\makecell{{\it Wave-like}\\ dispersion relation.}& \makecell{Non wave-like\\ dispersion relation.\\ {\it A ghost}\\ (in high momentum)\\ if the graviton\\ is healthy${}^{*}$.} & \makecell{{Non wave-like}\\ dispersion relation.} \\
\hline 
Graviton & \multicolumn{4}{c}{\makecell{ Is massless.\\ The no ghost, stability and subluminality  conditions $(\mathcal{G}_\tau>0,\, \mathcal{F}_\tau >0,\, \frac{\mathcal{F}_\tau}{\mathcal{G}_\tau}<1)$\\ are satisfied if\\ $G_{4}>-2\,X\,G_{4,X}>0$ .}  } \\
\hline 

Vector sector & \multicolumn{4}{c}{Non dynamical.} \\
\hline 
    \end{tabular}
\end{table*}

\subsection{Discussions and outlook}\label{sec discussions}
It is important to notice that the result in (\ref{eqn finalql concl}) showing only a graviton and a single scalar perturbation about the FLRW background is contrary to a naive expectation, because there are terms in the action (\ref{eqn G4Tlag concl}) that suggest a kinetic mixing of the Horndeski scalar with Torsion, as we explained at the beginning of section \ref{sec main}. For instance, there are terms where torsion couples to second covariant derivatives of the scalar. This opens the question whether there are hidden symmetries in the full theory, or accidental  symmetries for linear order perturbations on this specific background, such as in other theories with torsion \cite{Golovnev:2018wbh,Golovnev:2020nln, BeltranJimenez:2019nns} and with non-metricity \cite{BeltranJimenez:2019tme}. This could mean a strong coupling, evident in a discontinuity in the number of degrees of freedom between perturbative expansions about different backgrounds. A Hamiltonian analysis could shed light on this regard, but given the complexity of these theories other potentially simpler approaches would be to perform  higher order perturbations as in \cite{BeltranJimenez:2019nns} or to explore less symmetric backgrounds. However, the latter may need  a case by case analysis depending on the background (See for instance  \cite{BeltranJimenez:2019tme}). All in all, contrasting the evidence presented in this work to results following the latter approaches could be relevant to understand whether the theories in this work also suffer from strong coupling or not, and to assess the viability of these theories at least for applications in Cosmology where the FLRW background is of primary importance\footnote{We thank an anonymous referee who brought references \cite{Golovnev:2020nln, BeltranJimenez:2019nns} to our attention and pointed out similar results and discussions in other theories with torsion.}.
\section*{Acknowledgements}
Most of the findings in this work were the result of the close collaboration with Valery Rubakov who passed away on 19.10.2022 while this manuscript was being prepared.

The authors thank V. Volkova for the thoughtful reading of the manuscript. 

The work of S.M. on Sec. 4 of this paper has been supported by Russian Science Foundation grant 19-12-00393, while the part of work on Sec. 2 has been supported by the Foundation for the Advancement of Theoretical Physics and Mathematics "BASIS".

\section{Appendix}
\subsection{Equations of motion in Quartic Horndeski Cartan}\label{asec eqns}

For completeness, we show below the equations for the scalar and metric. The equation for contortion was shown in expression (\ref{eqn K}). 

\subsubsection{Equation for the scalar}
A generally covariant form of the equation for the scalar was written in equation (\ref{eqn G4ddk}) in the form ${\mathcal{E}}_{\phi} =0 $, with
\begin{dmath}
F(K,\, \,\tilde\nabla K;\, \,\tilde\nabla^2 \phi,\, \,\tilde\nabla \phi, \,\tilde R)\,\text{$=$}\, N +2 \tilde{\nabla}_\alpha G_{4,X}\left(\tilde{\nabla}^\mu T^\lambda\,_\mu\,^\alpha\, \tilde{\nabla}_\lambda\phi-T^{\lambda\mu\alpha}\,T^\rho\,_{\mu\lambda}\, \tilde{\nabla}_\rho\phi+3T^\lambda\,_\mu\,^\alpha\, \tilde{\nabla}_\lambda \tilde{\nabla}^\mu\phi \right)
+2G_{4,XX}\left[\tilde{\nabla}^\alpha\phi\, \tilde{\nabla}_\rho \tilde{\nabla}^\rho\phi\left(T^\lambda\,_{\beta\alpha}\, \tilde{\nabla}_\lambda \tilde{\nabla}^\beta\phi+ \tilde{\nabla}^\mu\left(T^\lambda\,_{\mu\alpha} \tilde{\nabla}_\lambda\phi\right)\right) -T^\lambda\,_{\nu\alpha}\, \tilde{\nabla}^\alpha\phi \tilde{\nabla}_\mu \tilde{\nabla}^\nu\phi \tilde{\nabla}_\lambda \tilde{\nabla}^\mu\phi\right]
+G_{4,X}\left[2\left(\tilde{\nabla}^\beta T^\lambda\,_{\alpha\beta}\, \tilde{\nabla}_\lambda \tilde{\nabla}^\alpha\phi-g^{\alpha\mu}T^\lambda\,_{\alpha\beta}\, \tilde{\nabla}^\beta\left(T^\gamma\,_{\lambda\mu} \tilde{\nabla}_\gamma\phi\right)\right)+2T^\lambda\,_{\mu\nu}\,\left(\tilde{R}_{\lambda\rho}\,^{\nu\mu}\, \tilde{\nabla}^\rho\phi-T^{\rho\nu\mu}\, \tilde{\nabla}_\rho \tilde{\nabla}_\lambda\phi\right)
+ \tilde{\nabla}_\lambda\phi\left(-T^{\alpha\nu\mu}\, \tilde{\nabla}_\alpha T^\lambda\,_{\mu\nu}+\tilde{R}^\lambda\,_\alpha\,^{\nu\mu}\,T^\alpha\,_{\mu\nu}\,-\tilde{R}^{\alpha\mu\nu}\,_\mu\,T^\lambda\,_{\alpha\nu}-\tilde{R}^{\alpha\nu}\,_\nu\,^\mu\,T^\lambda\,_{\mu\alpha}\right)
-\tilde{\nabla}^\sigma\phi\left(2 \tilde{\nabla}^\nu\left(-K^\mu\,_{\mu\rho}\,K^\rho\,_{\nu\sigma}+K^\rho\,_{\mu\nu}\,K^\mu\,_{\rho\sigma}\right)+ \tilde{\nabla}_\sigma\left(K^\rho\,_{\mu\nu}\,K^\nu\,_\rho\,^\mu\,+K^\rho\,_\mu\,^\mu\,K^\nu\,_{\nu\rho}\,\right)+2\left(K^\rho\,_{\nu\sigma}\,R_\rho\,^\nu\,-K^\nu\,_{\nu\rho}\,R_\sigma\,^\rho\,\right)\right)\right]\\
(-1 + \, c) \biggl(\,{G_{4,{\phi}X}} K_{\alpha }{}^{\gamma \zeta } (K_{\beta \gamma \zeta } -  K_{\beta \zeta \gamma }) \nabla^{\alpha }\phi \nabla^{\beta }\phi + \,{G_{4,{\phi}XX}} K_{\beta }{}^{\zeta \iota } (- K_{\gamma \zeta \iota } + K_{\gamma \iota \zeta }) \nabla_{\alpha }\phi \nabla^{\alpha }\phi \nabla^{\beta }\phi \nabla^{\gamma }\phi + \,{G_{4,XX}} \nabla^{\alpha }\phi \nabla^{\beta }\phi \Bigl(K_{\alpha }{}^{\zeta \iota } \bigl((- K_{\beta \zeta \iota } + K_{\beta \iota \zeta }) \nabla_{\gamma }\nabla^{\gamma }\phi + 2 (- \nabla_{\gamma }K_{\beta \zeta \iota } + \nabla_{\gamma }K_{\beta \iota \zeta }) \nabla^{\gamma }\phi \bigr) - 2 K_{\beta }{}^{\zeta \iota } (K_{\gamma \zeta \iota } -  K_{\gamma \iota \zeta }) (\nabla_{\alpha }\nabla^{\gamma }\phi + \nabla^{\gamma }\nabla_{\alpha }\phi)\Bigr) + 2 \,{G_{4,X}} \Bigl(K_{\alpha }{}^{\gamma \zeta } (K_{\beta \gamma \zeta } -  K_{\beta \zeta \gamma }) \nabla^{\beta }\nabla^{\alpha }\phi + \nabla^{\alpha }\phi \bigl(K^{\beta \gamma \zeta } (\nabla_{\zeta }K_{\alpha \beta \gamma } -  \nabla_{\zeta }K_{\alpha \gamma \beta }) + K_{\alpha }{}^{\beta \gamma } (\nabla_{\zeta }K_{\beta \gamma }{}^{\zeta } -  \nabla_{\zeta }K_{\gamma \beta }{}^{\zeta })\bigr)\Bigr) + \,{G_{4,XXX}} K_{\gamma }{}^{\iota \kappa } (K_{\zeta \iota \kappa } -  K_{\zeta \kappa \iota }) \nabla^{\alpha }\phi \nabla_{\beta }\nabla_{\alpha }\phi \nabla^{\beta }\phi \nabla^{\gamma }\phi \nabla^{\zeta }\phi \biggr)\,,
\label{eqn scalar f}
\end{dmath}
\begin{dmath}
N\,\text{=}\,\mathcal{L}_{4c,\phi}+\mathcal{L}_{4c,X} \tilde{\nabla}_\nu \tilde{\nabla}^\nu\phi+\left(\tilde{R}\, g_{\lambda\nu}-4\tilde{R}_{\lambda\nu}\right)\tilde{\nabla}^{\lambda}\phi\tilde{\nabla}^{\nu}G_{4,X}-2G_{4,X}\tilde{R}_{\lambda\beta}\tilde{\nabla}^{\beta}\tilde{\nabla}^{\lambda}\phi 
+\tilde{\nabla}_{\alpha}G_{4,XX}\left[\tilde{\nabla}^\alpha\phi\left(\tilde{\nabla}_\nu \tilde{\nabla}^\nu\phi\right)^2-2\tilde{\nabla}^\mu\phi \tilde{\nabla}^\alpha\tilde{\nabla}_\mu\phi \,\tilde{\nabla}_\nu \tilde{\nabla}^\nu\phi -\tilde{\nabla}^\alpha \phi \tilde{\nabla}^\mu\tilde{\nabla}^\nu \phi\tilde{\nabla}_{\mu}\tilde{\nabla}_\nu\phi +2\tilde{\nabla}^\mu\phi \tilde{\nabla}_\nu\tilde{\nabla}_\mu\phi \tilde{\nabla}^\nu\tilde{\nabla}^\alpha\phi \right] 
+2\tilde{\nabla}_\beta G_{4,X\phi}\left[\tilde{\nabla}^\beta\phi\, \tilde{\nabla}_\nu \tilde{\nabla}^\nu\phi-\tilde{\nabla}_\alpha\phi \tilde{\nabla}^\alpha \tilde{\nabla}^\beta\phi\right] +2G_{4,X\phi}\left[\left(\tilde{\nabla}_\nu \tilde{\nabla}^\nu\phi\right)^2-\tilde{\nabla}_\beta \tilde{\nabla}_\alpha\phi \tilde{\nabla}^\alpha \tilde{\nabla}^\beta\phi\right]+2G_{4,XX}\left[\tilde{\nabla}^\alpha\phi \tilde{\nabla}^\lambda\phi\left(\tilde{R}^\mu\,_{\lambda\sigma\alpha}\tilde{\nabla}_\mu \tilde{\nabla}^\sigma\phi-\tilde{R}_{\lambda\alpha}\, \tilde{\nabla}_\nu \tilde{\nabla}^\nu\phi\right)+\left(\tilde{\nabla}_\alpha \tilde{\nabla}_\mu\phi \tilde{\nabla}_\beta \tilde{\nabla}^\mu\phi \tilde{\nabla}^\alpha \tilde{\nabla}^\beta\phi-\tilde{\nabla}_\alpha \tilde{\nabla}_\beta\phi \tilde{\nabla}^\alpha \tilde{\nabla}^\beta\phi\, \tilde{\nabla}_\nu \tilde{\nabla}^\nu\phi\right)\right]\,.\label{eqn h4}
\end{dmath}
and for a function $G(\phi,X)$
\begin{eqnarray}
\tilde{\nabla}_\alpha G(\phi,X)=G_{,\phi} \tilde{\nabla}_\alpha\phi-G_{,X}g^{\mu\nu} \tilde{\nabla}_\alpha \tilde{\nabla}_\mu\phi \tilde{\nabla}_\nu\phi\,.
\end{eqnarray}
$N$ in expression (\ref{eqn h4}) can be used to write the standard Quartic Horndeski equation for the scalar on a {\it torsionless} spacetime as $N=0$ by taking $\tilde{\nabla}\rightarrow \nabla$ and $\tilde{R}\rightarrow R$.

\subsubsection{Equation of motion for metric}

A generally covariant form of the equation for the metric was written in equation (\ref{eqn fieldeoms}) in the form ${\mathcal{E}}_{g_{\mu\nu}} =0 $, with
\begin{dmath}
{\mathcal{E}}_{g_{\mu\nu}}\,\text{=}\,\tfrac{1}{2} \,{G_{4}} \Bigl(-2 \bigl(K_{\alpha }{}^{\mu }{}_{\beta } K^{\nu \alpha \beta } + K^{\mu \alpha \beta } (K_{\alpha }{}^{\nu }{}_{\beta } + K^{\nu }{}_{\beta \alpha }) -  K^{\mu \alpha }{}_{\alpha } K^{\nu \beta }{}_{\beta } + K_{\alpha }{}^{\beta }{}_{\beta } (K^{\mu \nu \alpha } + K^{\nu \mu \alpha }) + R^{\mu \nu }\bigr) + g^{\mu \nu } (K_{\alpha \gamma \beta } K^{\alpha \beta \gamma } + K^{\alpha }{}_{\alpha }{}^{\beta } K_{\beta }{}^{\gamma }{}_{\gamma } + R)\Bigr) + \,{G_{4,{\phi}{\phi}}} (- g^{\mu \nu } \nabla_{\alpha }\phi \nabla^{\alpha }\phi + \nabla^{\mu }\phi \nabla^{\nu }\phi) + \tfrac{1}{2} \,{G_{4,{\phi}}} \bigl(-2 g^{\mu \nu } \nabla_{\alpha }\nabla^{\alpha }\phi + 2 (g^{\mu \nu } K_{\alpha }{}^{\beta }{}_{\beta } + K^{\mu \nu }{}_{\alpha } + K^{\nu \mu }{}_{\alpha }) \nabla^{\alpha }\phi - 2 K^{\nu \alpha }{}_{\alpha } \nabla^{\mu }\phi + \nabla^{\mu }\nabla^{\nu }\phi - 2 K^{\mu \alpha }{}_{\alpha } \nabla^{\nu }\phi + \nabla^{\nu }\nabla^{\mu }\phi \bigr) + \tfrac{1}{2} \,{G_{4,{\phi}X}} \Bigl(- K^{\mu \nu }{}_{\beta } \nabla_{\alpha }\phi \nabla^{\alpha }\phi \nabla^{\beta }\phi -  K^{\nu \mu }{}_{\beta } \nabla_{\alpha }\phi \nabla^{\alpha }\phi \nabla^{\beta }\phi - 2 g^{\mu \nu } \nabla^{\alpha }\phi \bigl(-2 \nabla_{\beta }\nabla_{\alpha }\phi \nabla^{\beta }\phi + \nabla_{\alpha }\phi (\nabla_{\beta }\nabla^{\beta }\phi + K_{\beta }{}^{\gamma }{}_{\gamma } \nabla^{\beta }\phi)\bigr) -  \nabla_{\alpha }\nabla^{\nu }\phi \nabla^{\alpha }\phi \nabla^{\mu }\phi + K^{\nu }{}_{\alpha \beta } \nabla^{\alpha }\phi \nabla^{\beta }\phi \nabla^{\mu }\phi + \nabla_{\alpha }\phi \nabla^{\alpha }\phi \nabla^{\mu }\nabla^{\nu }\phi -  \nabla_{\alpha }\nabla^{\mu }\phi \nabla^{\alpha }\phi \nabla^{\nu }\phi + K^{\mu }{}_{\alpha \beta } \nabla^{\alpha }\phi \nabla^{\beta }\phi \nabla^{\nu }\phi + 4 \nabla_{\alpha }\nabla^{\alpha }\phi \nabla^{\mu }\phi \nabla^{\nu }\phi + 4 K_{\alpha }{}^{\beta }{}_{\beta } \nabla^{\alpha }\phi \nabla^{\mu }\phi \nabla^{\nu }\phi - 3 \nabla^{\alpha }\phi \nabla^{\mu }\nabla_{\alpha }\phi \nabla^{\nu }\phi - 3 \nabla^{\alpha }\phi \nabla^{\mu }\phi \nabla^{\nu }\nabla_{\alpha }\phi + \nabla_{\alpha }\phi \nabla^{\alpha }\phi \nabla^{\nu }\nabla^{\mu }\phi \Bigr) + \tfrac{1}{2} \,{G_{4,XX}} \Bigl(K^{\mu \nu }{}_{\gamma } \nabla^{\alpha }\phi \nabla_{\beta }\nabla_{\alpha }\phi \nabla^{\beta }\phi \nabla^{\gamma }\phi + K^{\nu \mu }{}_{\gamma } \nabla^{\alpha }\phi \nabla_{\beta }\nabla_{\alpha }\phi \nabla^{\beta }\phi \nabla^{\gamma }\phi + 2 g^{\mu \nu } \nabla^{\alpha }\phi \nabla^{\beta }\phi \bigl(\nabla_{\beta }\nabla_{\alpha }\phi (\nabla_{\gamma }\nabla^{\gamma }\phi + K_{\gamma }{}^{\zeta }{}_{\zeta } \nabla^{\gamma }\phi) -  \nabla_{\gamma }\nabla_{\beta }\phi \nabla^{\gamma }\nabla_{\alpha }\phi \bigr) + \nabla^{\alpha }\phi \nabla_{\beta }\nabla_{\alpha }\phi \nabla^{\beta }\nabla^{\nu }\phi \nabla^{\mu }\phi + K_{\beta }{}^{\nu }{}_{\gamma } \nabla^{\alpha }\phi \nabla^{\beta }\phi \nabla^{\gamma }\nabla_{\alpha }\phi \nabla^{\mu }\phi -  K^{\nu }{}_{\gamma \beta } \nabla^{\alpha }\phi \nabla^{\beta }\phi \nabla^{\gamma }\nabla_{\alpha }\phi \nabla^{\mu }\phi -  \nabla^{\alpha }\phi \nabla_{\beta }\nabla_{\alpha }\phi \nabla^{\beta }\phi \nabla^{\mu }\nabla^{\nu }\phi + \nabla^{\alpha }\phi \nabla_{\beta }\nabla_{\alpha }\phi \nabla^{\beta }\nabla^{\mu }\phi \nabla^{\nu }\phi + K_{\beta }{}^{\mu }{}_{\gamma } \nabla^{\alpha }\phi \nabla^{\beta }\phi \nabla^{\gamma }\nabla_{\alpha }\phi \nabla^{\nu }\phi -  K^{\mu }{}_{\gamma \beta } \nabla^{\alpha }\phi \nabla^{\beta }\phi \nabla^{\gamma }\nabla_{\alpha }\phi \nabla^{\nu }\phi + \nabla_{\alpha }\nabla^{\alpha }\phi \nabla_{\beta }\nabla^{\beta }\phi \nabla^{\mu }\phi \nabla^{\nu }\phi + 2 K_{\alpha }{}^{\gamma }{}_{\gamma } \nabla^{\alpha }\phi \nabla_{\beta }\nabla^{\beta }\phi \nabla^{\mu }\phi \nabla^{\nu }\phi -  K_{\alpha }{}^{\gamma \zeta } K_{\beta \zeta \gamma } \nabla^{\alpha }\phi \nabla^{\beta }\phi \nabla^{\mu }\phi \nabla^{\nu }\phi + K_{\alpha }{}^{\gamma }{}_{\gamma } K_{\beta }{}^{\zeta }{}_{\zeta } \nabla^{\alpha }\phi \nabla^{\beta }\phi \nabla^{\mu }\phi \nabla^{\nu }\phi -  \nabla_{\beta }\nabla_{\alpha }\phi \nabla^{\beta }\nabla^{\alpha }\phi \nabla^{\mu }\phi \nabla^{\nu }\phi - 2 K_{\alpha \gamma \beta } \nabla^{\alpha }\phi \nabla^{\gamma }\nabla^{\beta }\phi \nabla^{\mu }\phi \nabla^{\nu }\phi - 2 \nabla^{\alpha }\phi \nabla_{\beta }\nabla^{\beta }\phi \nabla^{\mu }\nabla_{\alpha }\phi \nabla^{\nu }\phi - 2 K_{\beta }{}^{\gamma }{}_{\gamma } \nabla^{\alpha }\phi \nabla^{\beta }\phi \nabla^{\mu }\nabla_{\alpha }\phi \nabla^{\nu }\phi + \nabla^{\alpha }\phi \nabla_{\beta }\nabla_{\alpha }\phi \nabla^{\mu }\nabla^{\beta }\phi \nabla^{\nu }\phi - 2 \nabla^{\alpha }\phi \nabla_{\beta }\nabla^{\beta }\phi \nabla^{\mu }\phi \nabla^{\nu }\nabla_{\alpha }\phi - 2 K_{\beta }{}^{\gamma }{}_{\gamma } \nabla^{\alpha }\phi \nabla^{\beta }\phi \nabla^{\mu }\phi \nabla^{\nu }\nabla_{\alpha }\phi + 2 \nabla^{\alpha }\phi \nabla^{\beta }\phi \nabla^{\mu }\nabla_{\alpha }\phi \nabla^{\nu }\nabla_{\beta }\phi + \nabla^{\alpha }\phi \nabla_{\beta }\nabla_{\alpha }\phi \nabla^{\mu }\phi \nabla^{\nu }\nabla^{\beta }\phi -  \nabla^{\alpha }\phi \nabla_{\beta }\nabla_{\alpha }\phi \nabla^{\beta }\phi \nabla^{\nu }\nabla^{\mu }\phi \Bigr)
\end{dmath}
\begin{dmath*}
+ \tfrac{1}{2} \,{G_{4,X}} \biggl(- K^{\nu \mu }{}_{\beta } \nabla_{\alpha }\nabla^{\beta }\phi \nabla^{\alpha }\phi + \nabla_{\alpha }\nabla^{\mu }\nabla^{\nu }\phi \nabla^{\alpha }\phi + \nabla_{\alpha }\nabla^{\nu }\nabla^{\mu }\phi \nabla^{\alpha }\phi + K^{\mu \nu }{}_{\alpha } \nabla^{\alpha }\phi \nabla_{\beta }\nabla^{\beta }\phi + K^{\nu \mu }{}_{\alpha } \nabla^{\alpha }\phi \nabla_{\beta }\nabla^{\beta }\phi - 2 K_{\beta }{}^{\nu }{}_{\gamma } K^{\mu \gamma }{}_{\alpha } \nabla^{\alpha }\phi \nabla^{\beta }\phi + 2 K_{\beta }{}^{\gamma }{}_{\gamma } K^{\mu \nu }{}_{\alpha } \nabla^{\alpha }\phi \nabla^{\beta }\phi - 2 K_{\beta }{}^{\mu }{}_{\gamma } K^{\nu \gamma }{}_{\alpha } \nabla^{\alpha }\phi \nabla^{\beta }\phi + 2 K_{\beta }{}^{\gamma }{}_{\gamma } K^{\nu \mu }{}_{\alpha } \nabla^{\alpha }\phi \nabla^{\beta }\phi -  \nabla^{\alpha }\phi \nabla_{\beta }K^{\mu \nu }{}_{\alpha } \nabla^{\beta }\phi -  \nabla^{\alpha }\phi \nabla_{\beta }K^{\nu \mu }{}_{\alpha } \nabla^{\beta }\phi - 2 K^{\nu \mu }{}_{\beta } \nabla^{\alpha }\phi \nabla^{\beta }\nabla_{\alpha }\phi -  K^{\mu \nu }{}_{\beta } \nabla^{\alpha }\phi (\nabla_{\alpha }\nabla^{\beta }\phi + 2 \nabla^{\beta }\nabla_{\alpha }\phi) -  K_{\alpha }{}^{\nu }{}_{\beta } \nabla^{\alpha }\phi \nabla^{\beta }\nabla^{\mu }\phi -  K^{\nu }{}_{\beta \alpha } \nabla^{\alpha }\phi \nabla^{\beta }\nabla^{\mu }\phi -  K_{\alpha }{}^{\mu }{}_{\beta } \nabla^{\alpha }\phi \nabla^{\beta }\nabla^{\nu }\phi -  K^{\mu }{}_{\beta \alpha } \nabla^{\alpha }\phi \nabla^{\beta }\nabla^{\nu }\phi 
-  g^{\mu \nu } \Bigl(\nabla_{\alpha }\nabla^{\alpha }\phi \nabla_{\beta }\nabla^{\beta }\phi \\-  \nabla_{\beta }\nabla_{\alpha }\phi \nabla^{\beta }\nabla^{\alpha }\phi + \nabla^{\alpha }\phi \bigl(2 \nabla_{\alpha }\nabla_{\beta }\nabla^{\beta }\phi - 2 \nabla_{\beta }\nabla^{\beta }\nabla_{\alpha }\phi + (K_{\alpha }{}^{\gamma \zeta } K_{\beta \zeta \gamma } -  K_{\alpha }{}^{\gamma }{}_{\gamma } K_{\beta }{}^{\zeta }{}_{\zeta } + 2 \nabla_{\beta }K_{\alpha }{}^{\gamma }{}_{\gamma }) \nabla^{\beta }\phi \\
+ 2 K_{\beta }{}^{\gamma }{}_{\gamma } (\nabla_{\alpha }\nabla^{\beta }\phi + \nabla^{\beta }\nabla_{\alpha }\phi) + 2 K_{\alpha \gamma \beta } \nabla^{\gamma }\nabla^{\beta }\phi \bigr)\Bigr) - 2 K^{\nu \beta }{}_{\beta } \nabla_{\alpha }\nabla^{\alpha }\phi \nabla^{\mu }\phi -  (\nabla_{\alpha }\nabla^{\alpha }\nabla^{\nu }\phi + \nabla_{\alpha }\nabla^{\nu }\nabla^{\alpha }\phi) \nabla^{\mu }\phi - 2 K_{\alpha }{}^{\gamma }{}_{\gamma } K^{\nu \beta }{}_{\beta } \nabla^{\alpha }\phi \nabla^{\mu }\phi + 2 K_{\alpha \gamma \beta } K^{\nu \beta \gamma } \nabla^{\alpha }\phi \nabla^{\mu }\phi -  \nabla^{\alpha }\phi \nabla_{\beta }K_{\alpha }{}^{\nu \beta } \nabla^{\mu }\phi + \nabla^{\alpha }\phi \nabla_{\beta }K^{\nu \beta }{}_{\alpha } \nabla^{\mu }\phi -  K_{\alpha }{}^{\nu }{}_{\beta } \nabla^{\beta }\nabla^{\alpha }\phi \nabla^{\mu }\phi + 3 K^{\nu }{}_{\beta \alpha } \nabla^{\beta }\nabla^{\alpha }\phi \nabla^{\mu }\phi + 2 K^{\nu \beta }{}_{\beta } \nabla^{\alpha }\phi \nabla^{\mu }\nabla_{\alpha }\phi -  \nabla^{\alpha }\nabla^{\nu }\phi \nabla^{\mu }\nabla_{\alpha }\phi + 2 K_{\alpha }{}^{\nu }{}_{\beta } \nabla^{\alpha }\phi \nabla^{\mu }\nabla^{\beta }\phi + \nabla_{\alpha }\nabla^{\alpha }\phi \nabla^{\mu }\nabla^{\nu }\phi -  \nabla^{\alpha }\phi \nabla^{\mu }\nabla^{\nu }\nabla_{\alpha }\phi + 2 \nabla^{\alpha }\phi \nabla^{\mu }\phi \nabla^{\nu }K_{\alpha }{}^{\beta }{}_{\beta } - 2 K^{\mu \beta }{}_{\beta } \nabla_{\alpha }\nabla^{\alpha }\phi \nabla^{\nu }\phi -  \nabla_{\alpha }\nabla^{\alpha }\nabla^{\mu }\phi \nabla^{\nu }\phi -  \nabla_{\alpha }\nabla^{\mu }\nabla^{\alpha }\phi \nabla^{\nu }\phi - 2 K_{\alpha }{}^{\gamma }{}_{\gamma } K^{\mu \beta }{}_{\beta } \nabla^{\alpha }\phi \nabla^{\nu }\phi + 2 K_{\alpha \gamma \beta } K^{\mu \beta \gamma } \nabla^{\alpha }\phi \nabla^{\nu }\phi -  \nabla^{\alpha }\phi \nabla_{\beta }K_{\alpha }{}^{\mu \beta } \nabla^{\nu }\phi + \nabla^{\alpha }\phi \nabla_{\beta }K^{\mu \beta }{}_{\alpha } \nabla^{\nu }\phi -  K_{\alpha }{}^{\mu }{}_{\beta } \nabla^{\beta }\nabla^{\alpha }\phi \nabla^{\nu }\phi + 3 K^{\mu }{}_{\beta \alpha } \nabla^{\beta }\nabla^{\alpha }\phi \nabla^{\nu }\phi + 2 \nabla^{\alpha }\phi \nabla^{\mu }K_{\alpha }{}^{\beta }{}_{\beta } \nabla^{\nu }\phi + K_{\alpha \gamma \beta } K^{\alpha \beta \gamma } \nabla^{\mu }\phi \nabla^{\nu }\phi + K^{\alpha }{}_{\alpha }{}^{\beta } K_{\beta }{}^{\gamma }{}_{\gamma } \nabla^{\mu }\phi \nabla^{\nu }\phi + R \,\nabla^{\mu }\phi \nabla^{\nu }\phi + 2 \nabla_{\beta }K^{\alpha }{}_{\alpha }{}^{\beta } \nabla^{\mu }\phi \nabla^{\nu }\phi + 2 \nabla^{\mu }\nabla_{\alpha }\nabla^{\alpha }\phi \nabla^{\nu }\phi + 2 K_{\alpha }{}^{\beta }{}_{\beta } \nabla^{\mu }\nabla^{\alpha }\phi \nabla^{\nu }\phi + 2 K^{\mu \beta }{}_{\beta } \nabla^{\alpha }\phi \nabla^{\nu }\nabla_{\alpha }\phi -  \nabla^{\alpha }\nabla^{\mu }\phi \nabla^{\nu }\nabla_{\alpha }\phi + 2 \nabla^{\mu }\phi \nabla^{\nu }\nabla_{\alpha }\nabla^{\alpha }\phi + 2 K_{\alpha }{}^{\beta }{}_{\beta } \nabla^{\mu }\phi \nabla^{\nu }\nabla^{\alpha }\phi + 2 K_{\alpha }{}^{\mu }{}_{\beta } \nabla^{\alpha }\phi \nabla^{\nu }\nabla^{\beta }\phi + \nabla_{\alpha }\nabla^{\alpha }\phi \nabla^{\nu }\nabla^{\mu }\phi -  \nabla^{\alpha }\phi \nabla^{\nu }\nabla^{\mu }\nabla_{\alpha }\phi \biggr)\\
+ \, c \Biggl(\tfrac{1}{2} \,{G_{4,XX}} K_{\alpha }{}^{\gamma \zeta } (- K_{\beta \gamma \zeta } + K_{\beta \zeta \gamma }) \nabla^{\alpha }\phi \nabla^{\beta }\phi \nabla^{\mu }\phi \nabla^{\nu }\phi + \tfrac{1}{2} \,{G_{4,X}} \nabla^{\alpha }\phi \biggl(\Bigl(g^{\mu \nu } K_{\alpha }{}^{\gamma \zeta } (- K_{\beta \gamma \zeta } + K_{\beta \zeta \gamma }) + 2 \bigl(K_{\beta }{}^{\nu }{}_{\gamma } (K_{\alpha }{}^{\mu \gamma } + K^{\mu \gamma }{}_{\alpha }) + K^{\mu \gamma }{}_{\alpha } K^{\nu }{}_{\gamma \beta } + K_{\beta }{}^{\mu }{}_{\gamma } K^{\nu \gamma }{}_{\alpha }\bigr)\Bigr) \nabla^{\beta }\phi + 2 (K_{\alpha \beta \gamma } -  K_{\alpha \gamma \beta }) (K^{\nu \beta \gamma } \nabla^{\mu }\phi + K^{\mu \beta \gamma } \nabla^{\nu }\phi)\biggr)\Biggr)\,,
\end{dmath*}
where higher covariant derivatives can be re expressed in terms of curvature tensors and lower derivatives using their commutator.



\subsection{Vector sector and coefficients of the quadratic action in its initial form and after using constraints}\label{asec ql0}

\paragraph{Tensor sector:}

 The coefficients $v_i$, $i=1,\dots 7$ in the action for the tensor sector (\ref{eqn ql0t}) are 

\begin{eqnarray}
v_{1}&=& 2 \left( \, a ^2\, \, G_{4} \,  -  \, \dot{\varphi} ^2\, \, G_{4,X} \, \right)\,,\\
v_{2}&=& - 2 \, a^2\, G_{4}\,,\\
v_{3}&=&-\frac{2}{a^4}\left( a^2 \,G_{4} +\, \, \dot{\varphi} ^2\, \, G_{4,X} \,\right)\,,\\
v_{4}&=& \frac{ 2\, G_{4}}{\, a^2}\,,\\
v_{5}&=& \,\left(\frac{8\, \, G_{4} \, \, \, x  }{\, a ^2} + \frac{8\,( \, x   + 2\, \, a \, \, \dot{a} )\, \, \dot{\varphi} ^2\, \, G_{4,X} \, }{ \, a ^4} - 8\, \, \dot{\varphi} \, \, G_{4,\phi} \, \right)\,,\\
v_6&=& \frac{4\,\dot{\varphi} ^2\, \, G_{4,X} }{a ^2}\,,
\end{eqnarray}
\begin{dmath}
v_7= \,\frac{4}{a^2}\, \, G_{4} \, \left(3\, \, x  ^2 +  a^2\, \left(\dot{a} ^2 - 2\, \, a \, \, \ddot{a} \right)\right) + \frac{4}{\, a ^4}\, G_{4,X} \,( \, \dot{\varphi}^2 \,(3\, \, x  ^2 +  \, a ^2\,(-3\, \, \dot{a} ^2 +  \, \dot{x}   + 2\, \, a \, \, \ddot{a} )) + 2\, \, a ^2\,( \, x   +  \, a \, \, \dot{a} )\, \, \ddot{\varphi}\, \dot{\varphi} ) + \frac{4}{\,  a ^5} G_{4,XX} (( \, x   + 2\, \, a \, \, \dot{a} )\, \, \dot{\varphi} ^3\,(- \, \dot{a} \, \, \dot{\varphi}  +  \, a \, \, \ddot{\varphi} ))\,  + 4\, G_{4,\phi}(-( \, x   +  \, a \, \, \dot{a} )\, \, \dot{\varphi}  -  \, a ^2\, \, \ddot{\varphi} ) - 4 \,a^2\,  \dot{\varphi} ^2\, \, G_{4,\phi\phi} + \frac{4}{a^2}\, G_{4,X\phi} \, \dot{\varphi} ^2\,\left(( \, x   + 3\, \, a \, \, \dot{a} )\, \, \dot{\varphi} -  a^2\, \ddot{\varphi} \right)\,\,.
\end{dmath} \bigskip

\paragraph{Vector sector:}

 The quadratic action for the vector perturbations can be written as

\begin{dmath}\mathcal{S}^{Vector}_{c}= \frac{1}{2}\,\int \text{d}\eta\, \text{d}^3x \,\left( h_{1} \, \,{{V^{\scalebox{0.5}{(1)}}_{j}}} \,{{V^{\scalebox{0.5}{(5)}}_{j}}} \, + \, h_{2} \, \,{{S_{j}}} \,{{V^{\scalebox{0.5}{(5)}}_{j}}} \, + \, h_{3} \, \,{\partial_{i}F_{j}} \,{\partial_{i}S_{j}} \, + \, h_{4} \, \,{{S_{j}}} \,{{V^{\scalebox{0.5}{(1)}}_{j}}} \, + \, h_{5} \, \,{\partial_{i}F_{j}} \,{\partial_{i}V^{\scalebox{0.5}{(2)}}_{j}} \, + \, h_{6} \, \,{\partial_{i}S_{j}} \,{\partial_{i}V^{\scalebox{0.5}{(2)}}_{j}} \, + \, h_{7} \, \,{\partial_{i}F_{j}} \,{\partial_{i}V^{\scalebox{0.5}{(3)}}_{j}} \, + \, h_{8} \, \,{\partial_{i}S_{j}} \,{\partial_{i}V^{\scalebox{0.5}{(3)}}_{j}} \, + \, h_{9} \, \,{\partial_{i}V^{\scalebox{0.5}{(2)}}_{j}} \,{\partial_{i}V^{\scalebox{0.5}{(3)}}_{j}} \, + \, h_{10} \, \,{\partial_{i}V^{\scalebox{0.5}{(2)}}_{j}} \,{\partial_{i}V^{\scalebox{0.5}{(4)}}_{j}} \, + \, h_{11} \, \,{\partial_{i}V^{\scalebox{0.5}{(3)}}_{j}} \,{\partial_{i}V^{\scalebox{0.5}{(4)}}_{j}} \, + \, h_{12} \, \,{\partial_{i}S_{j}} \,{\partial_{i}V^{\scalebox{0.5}{(6)}}_{j}} \, + \, h_{13} \, \,{\partial_{i}V^{\scalebox{0.5}{(1)}}_{j}} \,{\partial_{i}V^{\scalebox{0.5}{(6)}}_{j}} \, + \, h_{14} \, \,{\partial_{i}V^{\scalebox{0.5}{(5)}}_{j}} \,{\partial_{i}V^{\scalebox{0.5}{(6)}}_{j}} \, + \, h_{15} \, \,{\partial_{i}F_{j}} \,{\partial_{i}\dot{S_{j}}} \, + \, h_{16} \, \,{\partial_{i}\dot{F_{j}}} \,{\partial_{i}V^{\scalebox{0.5}{(2)}}_{j}} \, + \, h_{17} \, \,{\partial_{i}\dot{F_{j}}} \,{\partial_{i}V^{\scalebox{0.5}{(3)}}_{j}} \, + \, h_{18} \, \,({{V^{\scalebox{0.5}{(5)}}_{j}}})^2 \, + \, h_{19} \, \,({{S_{j}}})^2 \, + \, h_{20} \, \,({\partial_{i}S_{j}})^2 \, + \, h_{21} \, \,({\partial_{i}F_{j}})^2 \, + \, h_{22} \, \,({{V^{\scalebox{0.5}{(1)}}_{j}}})^2 \, + \, h_{23} \, \,({\partial_{i}V^{\scalebox{0.5}{(2)}}_{j}})^2 \, + \, h_{24} \, \,({\partial_{i}V^{\scalebox{0.5}{(3)}}_{j}})^2 \, + \, h_{25} \, \,({\partial_{i}\dot{F_{j}}})^2 \right)\,. \end{dmath}

Let us note that even though this action depends on $c$, the dynamics is independent of this parameter. Namely, there is no dynamical vector perturbation. The coefficients $h_i$, $i=1,\dots 25$ are

\begin{dmath}h_{1}=- \frac{8 \,{G_{4}}}{\,{a}^2} \,,\end{dmath}\begin{dmath}h_{2}=\frac{8 \,{a}^2 \,{G_{4}} \,{x} - 8 \,{a}^4 \,{G_{4,{\phi}}} \,\dot{\varphi} + 16 \,{G_{4,X}} (\,{x} + \,{a} \,\dot{a}) \,\dot{\varphi}^2}{\,{a}^4} \,,\end{dmath}\begin{dmath}h_{3}=\frac{4 \,{a}^4 \,{G_{4}} \,\dot{a} + 2 \,{a}^5 \,{G_{4,{\phi}}} \,\dot{\varphi} - 2 \,{a}^3 \,{G_{4,{\phi}X}} \,\dot{\varphi}^3 + 2 \,{G_{4,XX}} \,\dot{\varphi}^3 (- \,{a} \,\ddot{\varphi} + \,\dot{a} \,\dot{\varphi}) - 2 \,{a} \,{G_{4,X}} \,\dot{\varphi} \bigl(\,{a}^2 \,\ddot{\varphi} + (2 \,{x} + \,{a} \,\dot{a}) \,\dot{\varphi}\bigr)}{\,{a}^3} \,,\end{dmath}\begin{dmath}h_{4}=-\frac{2\,c\,x}{a^2}h_{6} \,,\end{dmath}\begin{dmath}h_{5}=\frac{4 \,{a}^2 \,{G_{4}} \,{x} - 4 \,{a}^4 \,{G_{4,{\phi}}} \,\dot{\varphi} + 4 \,{G_{4,X}} (\,{x} + 2 \,{a} \,\dot{a}) \,\dot{\varphi}^2}{\,{a}^4} \,,\end{dmath}\begin{dmath}h_{6}=- \frac{2 \,{G_{4,X}} \,\dot{\varphi}^2}{\,{a}^2} \,,\end{dmath}\begin{dmath}h_{7}=\frac{4 \,{a}^2 \,{G_{4}} \,{x} - 4 \,{a}^4 \,{G_{4,{\phi}}} \,\dot{\varphi} + 4 \,{G_{4,X}} (\,{x} + 2 \,{a} \,\dot{a}) \,\dot{\varphi}^2}{\,{a}^4} \,,\end{dmath}\begin{dmath}h_{8}=h_{6} \,,\end{dmath}\begin{dmath}h_{9}=- \frac{4 \,{a}^2 \,{G_{4}} - 4 (-1 + \, c) \,{G_{4,X}} \,\dot{\varphi}^2}{\,{a}^4} \,,\end{dmath}\begin{dmath}h_{10}=\frac{1}{2}h_{1} \,,\end{dmath}\begin{dmath}h_{11}=-\frac{1}{2}h_{1} \,,\end{dmath}\begin{dmath}h_{12}=- \frac{4 \,{a}^2 \,{G_{4}} \,{x} - 4 \,{a}^4 \,{G_{4,{\phi}}} \,\dot{\varphi} + 8 \,{G_{4,X}} (\,{x} + \,{a} \,\dot{a}) \,\dot{\varphi}^2}{\,{a}^4} \,,\end{dmath}\begin{dmath}h_{13}= -\frac{1}{2}h_{1} \,,\end{dmath}\begin{dmath}h_{14}= -\frac{1}{2}h_{1} \,,\end{dmath}\begin{dmath}h_{15}=2 \,{a}^2 \,{G_{4}} - 2 \,{G_{4,X}} \,\dot{\varphi}^2 \,,\end{dmath}\begin{dmath}h_{16}=-h_{6} \,,\end{dmath}\begin{dmath}h_{17}=-h_{6} \,,\end{dmath}\begin{dmath}h_{18}= \frac{1}{2}h_{1} \,,\end{dmath}\begin{dmath}h_{19}=\frac{1 }{\,{a}^6}\left(\,{G_{4}} (-6 \,{a}^4 \,{x}^2 + 6 \,{a}^6 \,\dot{a}^2) + 6 \,{a}^6 \,{G_{4,{\phi}}} (\,{x} + \,{a} \,\dot{a}) \,\dot{\varphi} - 2 \,{a}^2 \,{G_{4,X}} \bigl(- (-12 + \, c) \,{x}^2 + 18 \,{a} \,{x} \,\dot{a} + 6 \,{a}^2 \,\dot{a}^2\bigr) \,\dot{\varphi}^2 + 6 \,{a}^4 \,{G_{4,{\phi}X}} (\,{x} + \,{a} \,\dot{a}) \,\dot{\varphi}^3 - 6 \,{G_{4,XX}} (\,{x} + \,{a} \,\dot{a})^2 \,\dot{\varphi}^4\right) \,,\end{dmath}\begin{dmath}h_{20}=\,{a}^2 \,{G_{4}} -  \,{G_{4,X}} \,\dot{\varphi}^2 \,,\end{dmath}\begin{dmath}h_{21}=\frac{1}{\,{a}^5}\left(\,{G_{4}} (6 \,{a}^3 \,{x}^2 - 4 \,{a}^6 \,\ddot{a} + 2 \,{a}^5 \,\dot{a}^2) - 2 \,{a}^7 \,{G_{4,{\phi}{\phi}}} \,\dot{\varphi}^2 + 2 \,{G_{4,XX}} (\,{x} + 2 \,{a} \,\dot{a}) \,\dot{\varphi}^3 (\,{a} \,\ddot{\varphi} -  \,\dot{a} \,\dot{\varphi}) + \,{G_{4,{\phi}}} \bigl(-2 \,{a}^7 \,\ddot{\varphi} - 2 \,{a}^5 (\,{x} + \,{a} \,\dot{a}) \,\dot{\varphi}\bigr) + 2 \,{a}^3 \,{G_{4,{\phi}X}} \,\dot{\varphi}^2 \bigl(- \,{a}^2 \,\ddot{\varphi} + (\,{x} + 3 \,{a} \,\dot{a}) \,\dot{\varphi}\bigr) + 2 \,{a} \,{G_{4,X}} \,\dot{\varphi} \Bigl(2 \,{a}^2 \,\ddot{\varphi} (\,{x} + \,{a} \,\dot{a}) + \,\dot{\varphi} \bigl(3 \,{x}^2 + \,{a}^2 (2 \,{a} \,\ddot{a} - 3 \,\dot{a}^2 + \,\dot{x})\bigr)\Bigr)\right) \,,\end{dmath}\begin{dmath}h_{22}=-\frac{c}{a^2}h_{6} \,,\end{dmath}\begin{dmath}h_{23}=\frac{c}{a^2}h_{6} \,,\end{dmath}\begin{dmath}h_{24}=\frac{c}{a^2}h_{6} \,,\end{dmath}\begin{dmath}h_{25}=\,{a}^2 \,{G_{4}} -  \,{G_{4,X}} \,\dot{\varphi}^2 \,,\end{dmath} \bigskip

\paragraph{Scalar sector: }

 The coefficients $f_i$, $i=1,\dots 59$ in the action for the scalar sector (\ref{eqn ql0s}) are 

\begin{dmath}f_{1}= (x+\,a\,\dot{a})\left(-12 \,a^2\,{G_{4,{\phi}}} (x-\,a\,\dot{a}) + 12 \,{G_{4,{\phi}{\phi}}} \,\dot{\varphi} -  \frac{24 \,{G_{4,{\phi}X}} (2 \,{x} + \,{a} \,\dot{a}) \,\dot{\varphi}^2}{\,{a}^4} + \frac{12 \,{G_{4,{\phi}{\phi}X}} \,\dot{\varphi}^3}{\,{a}^2} -  \frac{12 \,{G_{4,{\phi}XX}} (\,{x} + \,{a} \,\dot{a}) \,\dot{\varphi}^4}{\,{a}^6}\right) \,,\end{dmath}\begin{dmath}f_{2}=- \frac{24 \,{G_{4}} \,{x}}{\,{a}^2} - 12 \,{G_{4,{\phi}}} \,\dot{\varphi} + \frac{24 \,{G_{4,X}} (- \,{x} + \,{a} \,\dot{a}) \,\dot{\varphi}^2}{\,{a}^4} \,,\end{dmath}\begin{dmath}f_{3}=\,{G_{4}} (- \frac{12 \,{x}^2}{\,{a}^2} - 36 \,\dot{a}^2) - 12 \,{G_{4,{\phi}}} (\,{x} + 3 \,{a} \,\dot{a}) \,\dot{\varphi} + \frac{24 \,{G_{4,X}} (-2 \,{x}^2 + 3 \,{a} \,{x} \,\dot{a} + 3 \,{a}^2 \,\dot{a}^2) \,\dot{\varphi}^2}{\,{a}^4} -  \frac{12 \,{G_{4,{\phi}X}} (\,{x} + 3 \,{a} \,\dot{a}) \,\dot{\varphi}^3}{\,{a}^2} -  \frac{12 \,{G_{4,XX}} (\,{x} - 3 \,{a} \,\dot{a}) (\,{x} + \,{a} \,\dot{a}) \,\dot{\varphi}^4}{\,{a}^6} \,,\end{dmath}\begin{dmath}f_{4}=12 \,{G_{4,{\phi}}} (\frac{\,{x}^2}{\,{a}^2} - 2 \,{a} \,\ddot{a} + \,\dot{a}^2) - 12 \,{a}^2 \,{G_{4,{\phi}{\phi}{\phi}}} \,\dot{\varphi}^2 + \frac{24 \,{G_{4,{\phi}XX}} (\,{x} + \,{a} \,\dot{a}) \,\dot{\varphi}^3 (\,{a} \,\ddot{\varphi} -  \,\dot{a} \,\dot{\varphi})}{\,{a}^5} + \,{G_{4,{\phi}{\phi}}} \bigl(-12 \,{a}^2 \,\ddot{\varphi} + 12 (\,{x} -  \,{a} \,\dot{a}) \,\dot{\varphi}\bigr) + 12 \,{G_{4,{\phi}{\phi}X}} \,\dot{\varphi}^2 \bigl(- \,\ddot{\varphi} + \frac{(2 \,{x} + 3 \,{a} \,\dot{a}) \,\dot{\varphi}}{\,{a}^2}\bigr) + \frac{12 \,{G_{4,{\phi}X}} \,\dot{\varphi} \biggl(2 \,{a}^2 \,\ddot{\varphi} (2 \,{x} + \,{a} \,\dot{a}) + \,\dot{\varphi} \Bigl(\,{x}^2 - 6 \,{a} \,{x} \,\dot{a} + \,{a}^2 \bigl(-3 \,\dot{a}^2 + 2 (\,{a} \,\ddot{a} + \,\dot{x})\bigr)\Bigr)\biggr)}{\,{a}^4} \,,\end{dmath}\begin{dmath}f_{5}=- \frac{24 \,{G_{4,{\phi}}} \,{x}}{\,{a}^2} + 12 \,{G_{4,{\phi}{\phi}}} \,\dot{\varphi} -  \frac{24 \,{G_{4,{\phi}X}} (\,{x} + \,{a} \,\dot{a}) \,\dot{\varphi}^2}{\,{a}^4} \,,\end{dmath}\begin{dmath}f_{6}=\frac{24 \,{G_{4}} \,{x}}{\,{a}^2} - 12 \,{G_{4,{\phi}}} \,\dot{\varphi} + \frac{24 \,{G_{4,X}} (4 \,{x} + 3 \,{a} \,\dot{a}) \,\dot{\varphi}^2}{\,{a}^4} -  \frac{12 \,{G_{4,{\phi}X}} \,\dot{\varphi}^3}{\,{a}^2} + \frac{24 \,{G_{4,XX}} (\,{x} + \,{a} \,\dot{a}) \,\dot{\varphi}^4}{\,{a}^6} \,,\end{dmath}\begin{dmath}f_{7}=\frac{4  \,{G_{4,X}} \,{x} \,\dot{\varphi}^2}{\,{a}^4} \,,\end{dmath}\begin{dmath}f_{8}=- \frac{8 \,{G_{4}} \,{x}}{\,{a}^2} - 4 \,{G_{4,{\phi}}} \,\dot{\varphi} + \frac{8 \,{G_{4,X}} (- \,{x} + \,{a} \,\dot{a}) \,\dot{\varphi}^2}{\,{a}^4} \,,\end{dmath}\begin{dmath}f_{9}=-\frac{2\,a^2}{x}\, f_7 \,,\end{dmath}\begin{dmath}f_{10}=f_{11}\,+\,\frac{1}{a^2}\,f_9 \,,\end{dmath}\begin{dmath}f_{11}=- \frac{8 \,{G_{4}}}{\,{a}^2} \,,\end{dmath}\begin{dmath}f_{12}=\frac{8 \,{G_{4}} \,{x}}{\,{a}^2} - 8 \,{G_{4,{\phi}}} \,\dot{\varphi} + \frac{16 \,{G_{4,X}} (\,{x} + \,{a} \,\dot{a}) \,\dot{\varphi}^2}{\,{a}^4} \,,\end{dmath}\begin{dmath}f_{13}=f_{11} \,,\end{dmath}\begin{dmath}f_{14}=\,{G_{4,{\phi}}} (8 \,{x} + 4 \,{a} \,\dot{a}) - 4 \,{a}^2 \,{G_{4,{\phi}{\phi}}} \,\dot{\varphi} + \frac{4 \,{G_{4,X}} \bigl((-7 + \, c) \,{x}^2 - 10 \,{a} \,{x} \,\dot{a} - 3 \,{a}^2 \,\dot{a}^2\bigr) \,\dot{\varphi}}{\,{a}^4} + \frac{20 \,{G_{4,{\phi}X}} (\,{x} + \,{a} \,\dot{a}) \,\dot{\varphi}^2}{\,{a}^2} -  \frac{12 \,{G_{4,XX}} (\,{x} + \,{a} \,\dot{a})^2 \,\dot{\varphi}^3}{\,{a}^6} \,,\end{dmath}\begin{dmath}f_{15}=4 \,{G_{4,{\phi}}} (\frac{\,{x}^2}{\,{a}^2} - 2 \,{a} \,\ddot{a} + \,\dot{a}^2) - 4 \,{a}^2 \,{G_{4,{\phi}{\phi}{\phi}}} \,\dot{\varphi}^2 + \frac{8 \,{G_{4,{\phi}XX}} (\,{x} + \,{a} \,\dot{a}) \,\dot{\varphi}^3 (\,{a} \,\ddot{\varphi} -  \,\dot{a} \,\dot{\varphi})}{\,{a}^5} + \,{G_{4,{\phi}{\phi}}} \bigl(-4 \,{a}^2 \,\ddot{\varphi} + 4 (\,{x} -  \,{a} \,\dot{a}) \,\dot{\varphi}\bigr) + 4 \,{G_{4,{\phi}{\phi}X}} \,\dot{\varphi}^2 \bigl(- \,\ddot{\varphi} + \frac{(2 \,{x} + 3 \,{a} \,\dot{a}) \,\dot{\varphi}}{\,{a}^2}\bigr) + \frac{4 \,{G_{4,{\phi}X}} \,\dot{\varphi} \biggl(2 \,{a}^2 \,\ddot{\varphi} (2 \,{x} + \,{a} \,\dot{a}) + \,\dot{\varphi} \Bigl(\,{x}^2 - 6 \,{a} \,{x} \,\dot{a} + \,{a}^2 \bigl(-3 \,\dot{a}^2 + 2 (\,{a} \,\ddot{a} + \,\dot{x})\bigr)\Bigr)\biggr)}{\,{a}^4} \,,\end{dmath}\begin{dmath}f_{16}=-4 \,{G_{4,{\phi}}} + \frac{4 \,{G_{4,X}} \bigl((2 + \, c) \,{x} + 2 \,{a} \,\dot{a}\bigr) \,\dot{\varphi}}{\,{a}^4} \,,\end{dmath}\begin{dmath}f_{17}=\frac{8 \,{G_{4,{\phi}}} \,{x}}{\,{a}^2} - 4 \,{G_{4,{\phi}{\phi}}} \,\dot{\varphi} + \frac{8 \,{G_{4,{\phi}X}} (\,{x} + \,{a} \,\dot{a}) \,\dot{\varphi}^2}{\,{a}^4} \,,\end{dmath}\begin{dmath}f_{18}= -\frac{2\,a^2}{x\,\dot{\varphi}}\,f_{7} \,,\end{dmath}\begin{dmath}f_{19}=-8 \,{G_{4,{\phi}}} + \frac{\,{G_{4,X}} \bigl(8 \,{a}^2 \,\ddot{\varphi} + 8 (2 \,{x} + \,{a} \,\dot{a}) \,\dot{\varphi}\bigr)}{\,{a}^4} \,,\end{dmath}\begin{dmath}f_{20}=8 \,{a} \,{G_{4}} \,\dot{a} + 4 \,{a}^2 \,{G_{4,{\phi}}} \,\dot{\varphi} -  \frac{8 \,{G_{4,X}} (3 \,{x} + 2 \,{a} \,\dot{a}) \,\dot{\varphi}^2}{\,{a}^2} + 4 \,{G_{4,{\phi}X}} \,\dot{\varphi}^3 -  \frac{8 \,{G_{4,XX}} (\,{x} + \,{a} \,\dot{a}) \,\dot{\varphi}^4}{\,{a}^4} \,,\end{dmath}\begin{dmath}f_{21}=\,{G_{4}} (- \frac{4 \,{x}^2}{\,{a}^2} - 12 \,\dot{a}^2) - 4 \,{G_{4,{\phi}}} (\,{x} + 3 \,{a} \,\dot{a}) \,\dot{\varphi} + \frac{8 \,{G_{4,X}} (-2 \,{x}^2 + 3 \,{a} \,{x} \,\dot{a} + 3 \,{a}^2 \,\dot{a}^2) \,\dot{\varphi}^2}{\,{a}^4} -  \frac{4 \,{G_{4,{\phi}X}} (\,{x} + 3 \,{a} \,\dot{a}) \,\dot{\varphi}^3}{\,{a}^2} -  \frac{4 \,{G_{4,XX}} (\,{x} - 3 \,{a} \,\dot{a}) (\,{x} + \,{a} \,\dot{a}) \,\dot{\varphi}^4}{\,{a}^6} \,,\end{dmath}\begin{dmath}f_{22}=- \frac{8 \,{G_{4}} \,{x}}{\,{a}^2} + 4 \,{G_{4,{\phi}}} \,\dot{\varphi} -  \frac{8 \,{G_{4,X}} (4 \,{x} + 3 \,{a} \,\dot{a}) \,\dot{\varphi}^2}{\,{a}^4} + \frac{4 \,{G_{4,{\phi}X}} \,\dot{\varphi}^3}{\,{a}^2} -  \frac{8 \,{G_{4,XX}} (\,{x} + \,{a} \,\dot{a}) \,\dot{\varphi}^4}{\,{a}^6} \,,\end{dmath}\begin{dmath}f_{23}=\frac{2\,a^2}{x}\,f_{7} \,,\end{dmath}\begin{dmath}f_{24}=4 \,{a}^2 \,{G_{4,{\phi}}} -  \frac{8 \,{G_{4,X}} (2 \,{x} + \,{a} \,\dot{a}) \,\dot{\varphi}}{\,{a}^2} + 4 \,{G_{4,{\phi}X}} \,\dot{\varphi}^2 -  \frac{8 \,{G_{4,XX}} (\,{x} + \,{a} \,\dot{a}) \,\dot{\varphi}^3}{\,{a}^4} \,,\end{dmath}\begin{dmath}f_{25}=\,{G_{4}} (\frac{12 \,{x}^2}{\,{a}^2} + 8 \,{a} \,\ddot{a} - 4 \,\dot{a}^2) + 4 \,{a}^2 \,{G_{4,{\phi}{\phi}}} \,\dot{\varphi}^2 + \frac{8 \,{G_{4,XX}} (\,{x} -  \,{a} \,\dot{a}) \,\dot{\varphi}^3 (\,{a} \,\ddot{\varphi} -  \,\dot{a} \,\dot{\varphi})}{\,{a}^5} + 4 \,{G_{4,{\phi}X}} \,\dot{\varphi}^2 \bigl(\,\ddot{\varphi} + \frac{(2 \,{x} - 3 \,{a} \,\dot{a}) \,\dot{\varphi}}{\,{a}^2}\bigr) + \,{G_{4,{\phi}}} \bigl(4 \,{a}^2 \,\ddot{\varphi} + 4 (\,{x} + \,{a} \,\dot{a}) \,\dot{\varphi}\bigr) + \frac{4 \,{G_{4,X}} \,\dot{\varphi} \Bigl(-2 \,{a}^2 \,\ddot{\varphi} (-2 \,{x} + \,{a} \,\dot{a}) + \,\dot{\varphi} \bigl(3 \,{x}^2 - 6 \,{a} \,{x} \,\dot{a} + \,{a}^2 (-2 \,{a} \,\ddot{a} + 3 \,\dot{a}^2 + 2 \,\dot{x})\bigr)\Bigr)}{\,{a}^4} \,,\end{dmath}\begin{dmath}f_{26}= 4\, a^2\, f_{7} \,,\end{dmath}\begin{dmath}f_{27}=\frac{8 \,{G_{4}} \,{x}}{\,{a}^2} + 4 \,{G_{4,{\phi}}} \,\dot{\varphi} + \frac{8 \,{G_{4,X}} (\,{x} -  \,{a} \,\dot{a}) \,\dot{\varphi}^2}{\,{a}^4} \,,\end{dmath}\begin{dmath}f_{28}=-8 \,{a}^2 \,{G_{4,{\phi}}} + 8 \,{G_{4,{\phi}X}} \,\dot{\varphi}^2 + 8 \,{G_{4,X}} (\,\ddot{\varphi} + \frac{2 \,{x} \,\dot{\varphi}}{\,{a}^2}) + \frac{8 \,{G_{4,XX}} \,\dot{\varphi}^2 (\,{a} \,\ddot{\varphi} -  \,\dot{a} \,\dot{\varphi})}{\,{a}^3} \,,\end{dmath}\begin{dmath}f_{29}=-8 \,{a}^2 \,{G_{4}} + 8 \,{G_{4,X}} \,\dot{\varphi}^2 \,,\end{dmath}\begin{dmath}f_{30}=-f_{11} \,,\end{dmath}\begin{dmath}f_{31}=12 \,{G_{4,{\phi}}} (\,{x} + \,{a} \,\dot{a}) - 12 \,{a}^2 \,{G_{4,{\phi}{\phi}}} \,\dot{\varphi} -  \frac{12 \,{G_{4,X}} (-3 \,{x}^2 + 4 \,{a} \,{x} \,\dot{a} + 3 \,{a}^2 \,\dot{a}^2) \,\dot{\varphi}}{\,{a}^4} + \frac{12 \,{G_{4,{\phi}X}} (3 \,{x} + 5 \,{a} \,\dot{a}) \,\dot{\varphi}^2}{\,{a}^2} + \frac{12 \,{G_{4,XX}} (\,{x} - 3 \,{a} \,\dot{a}) (\,{x} + \,{a} \,\dot{a}) \,\dot{\varphi}^3}{\,{a}^6} \,,\end{dmath}\begin{dmath}f_{32}= (\,{x} + \,{a} \,\dot{a})\left(12 \,{G_{4,{\phi}}} -  \frac{36 \,{G_{4,X}} (3 \,{x} + \,{a} \,\dot{a}) \,\dot{\varphi}}{\,{a}^4} + \frac{48 \,{G_{4,{\phi}X}} \,\dot{\varphi}^2}{\,{a}^2} -  \frac{24 \,{G_{4,XX}} (4 \,{x} + 3 \,{a} \,\dot{a}) \,\dot{\varphi}^3}{\,{a}^6} + \frac{12 \,{G_{4,{\phi}XX}} \,\dot{\varphi}^4}{\,{a}^4} -  \frac{12 \,{G_{4,XXX}} \,\dot{\varphi}^5}{\,{a}^8} \right)\,,\end{dmath}\begin{dmath}f_{33}=12 \,{G_{4,{\phi}}} -  \frac{24 \,{G_{4,X}} (3 \,{x} + 2 \,{a} \,\dot{a}) \,\dot{\varphi}}{\,{a}^4} + \frac{12 \,{G_{4,{\phi}X}} \,\dot{\varphi}^2}{\,{a}^2} -  \frac{24 \,{G_{4,XX}} (\,{x} + \,{a} \,\dot{a}) \,\dot{\varphi}^3}{\,{a}^6} \,,\end{dmath}\begin{dmath}f_{34}=-24 \,{a} \,{G_{4}} \,\dot{a} - 12 \,{a}^2 \,{G_{4,{\phi}}} \,\dot{\varphi} + \frac{24 \,{G_{4,X}} (3 \,{x} + 2 \,{a} \,\dot{a}) \,\dot{\varphi}^2}{\,{a}^2} - 12 \,{G_{4,{\phi}X}} \,\dot{\varphi}^3 + \frac{24 \,{G_{4,XX}} (\,{x} + \,{a} \,\dot{a}) \,\dot{\varphi}^4}{\,{a}^4} \,,\end{dmath}\begin{dmath}f_{35}=\frac{6\, a^2}{x}\, f_{7} \,,\end{dmath}\begin{dmath}f_{36}=12 \,{a}^2 \,{G_{4,{\phi}}} -  \frac{24 \,{G_{4,X}} (2 \,{x} + \,{a} \,\dot{a}) \,\dot{\varphi}}{\,{a}^2} + 12 \,{G_{4,{\phi}X}} \,\dot{\varphi}^2 -  \frac{24 \,{G_{4,XX}} (\,{x} + \,{a} \,\dot{a}) \,\dot{\varphi}^3}{\,{a}^4} \,,\end{dmath}\begin{dmath}f_{37}= \frac{2\, a^2}{x} f_7 \,,\end{dmath}\begin{dmath}f_{38}=-8 \,{a} \,{G_{4}} \,\dot{a} - 4 \,{a}^2 \,{G_{4,{\phi}}} \,\dot{\varphi} + \frac{8 \,{G_{4,X}} (3 \,{x} + 2 \,{a} \,\dot{a}) \,\dot{\varphi}^2}{\,{a}^2} - 4 \,{G_{4,{\phi}X}} \,\dot{\varphi}^3 + \frac{8 \,{G_{4,XX}} (\,{x} + \,{a} \,\dot{a}) \,\dot{\varphi}^4}{\,{a}^4} \,,\end{dmath}\begin{dmath}f_{39}=-4 \,{a}^2 \,{G_{4,{\phi}}} + \frac{8 \,{G_{4,X}} (2 \,{x} + \,{a} \,\dot{a}) \,\dot{\varphi}}{\,{a}^2} - 4 \,{G_{4,{\phi}X}} \,\dot{\varphi}^2 + \frac{8 \,{G_{4,XX}} (\,{x} + \,{a} \,\dot{a}) \,\dot{\varphi}^3}{\,{a}^4} \,,\end{dmath}\begin{dmath}f_{40}=4 \,{G_{4,{\phi}}} (\,{x} + \,{a} \,\dot{a}) - 4 \,{a}^2 \,{G_{4,{\phi}{\phi}}} \,\dot{\varphi} -  \frac{4 \,{G_{4,X}} (-3 \,{x}^2 + 4 \,{a} \,{x} \,\dot{a} + 3 \,{a}^2 \,\dot{a}^2) \,\dot{\varphi}}{\,{a}^4} + \frac{4 \,{G_{4,{\phi}X}} (3 \,{x} + 5 \,{a} \,\dot{a}) \,\dot{\varphi}^2}{\,{a}^2} + \frac{4 \,{G_{4,XX}} (\,{x} - 3 \,{a} \,\dot{a}) (\,{x} + \,{a} \,\dot{a}) \,\dot{\varphi}^3}{\,{a}^6} \,,\end{dmath}\begin{dmath}f_{41}=-4 \,{G_{4,{\phi}}} + \frac{8 \,{G_{4,X}} (3 \,{x} + 2 \,{a} \,\dot{a}) \,\dot{\varphi}}{\,{a}^4} -  \frac{4 \,{G_{4,{\phi}X}} \,\dot{\varphi}^2}{\,{a}^2} + \frac{8 \,{G_{4,XX}} (\,{x} + \,{a} \,\dot{a}) \,\dot{\varphi}^3}{\,{a}^6} \,,\end{dmath}\begin{dmath}f_{42}=-\frac{2\,a^2}{x\dot{\varphi}}\,f_{7} \,,\end{dmath}\begin{dmath}f_{43}=4 \,{a}^2 \,{G_{4,{\phi}}} -  \frac{8 \,{G_{4,X}} (2 \,{x} + \,{a} \,\dot{a}) \,\dot{\varphi}}{\,{a}^2} + 4 \,{G_{4,{\phi}X}} \,\dot{\varphi}^2 -  \frac{8 \,{G_{4,XX}} (\,{x} + \,{a} \,\dot{a}) \,\dot{\varphi}^3}{\,{a}^4} \,,\end{dmath}\begin{dmath}f_{44}=8 \,{a}^2 \,{G_{4}} - 8 \,{G_{4,X}} \,\dot{\varphi}^2 \,,\end{dmath}\begin{dmath}f_{45}=-\frac{2\,a^2}{x}\, f_7 \,,\end{dmath}\begin{dmath}f_{46}=-8 \,{a}^2 \,{G_{4}} + 8 \,{G_{4,X}} \,\dot{\varphi}^2 \,,\end{dmath}\begin{dmath}f_{47}=\frac{12 \,{G_{4}}}{\,{a}^2} + \frac{12 \,{G_{4,X}} \,\dot{\varphi}^2}{\,{a}^4} \,,\end{dmath}\begin{dmath}f_{48}=\, (x +\,a\,\dot{a})\left(\frac{18\,G_{4}}{a^2} \,(x -\,a\,\dot{a}) - 18 \,{G_{4,{\phi}}} \,\dot{\varphi} + \frac{18 \,{G_{4,X}} (7 \,{x} + 3 \,{a} \,\dot{a}) \,\dot{\varphi}^2}{\,{a}^4} -  \frac{36 \,{G_{4,{\phi}X}} \,\dot{\varphi}^3}{\,{a}^2} + \frac{6 \,{G_{4,XX}} (11 \,{x} + 9 \,{a} \,\dot{a}) \,\dot{\varphi}^4}{\,{a}^6} -  \frac{6 \,{G_{4,{\phi}XX}} \,\dot{\varphi}^5}{\,{a}^4} + \frac{6 \,{G_{4,XXX}} (\,{x} + \,{a} \,\dot{a}) \,\dot{\varphi}^6}{\,{a}^8}\right) \,,\end{dmath}\begin{dmath}f_{49}=\,{G_{4}} (\frac{18 \,{x}^2}{\,{a}^2} + 12 \,{a} \,\ddot{a} - 6 \,\dot{a}^2) + 6 \,{a}^2 \,{G_{4,{\phi}{\phi}}} \,\dot{\varphi}^2 + \frac{12 \,{G_{4,XX}} (\,{x} -  \,{a} \,\dot{a}) \,\dot{\varphi}^3 (\,{a} \,\ddot{\varphi} -  \,\dot{a} \,\dot{\varphi})}{\,{a}^5} + 6 \,{G_{4,{\phi}X}} \,\dot{\varphi}^2 \bigl(\,\ddot{\varphi} + \frac{(2 \,{x} - 3 \,{a} \,\dot{a}) \,\dot{\varphi}}{\,{a}^2}\bigr) + \,{G_{4,{\phi}}} \bigl(6 \,{a}^2 \,\ddot{\varphi} + 6 (\,{x} + \,{a} \,\dot{a}) \,\dot{\varphi}\bigr) + \frac{6 \,{G_{4,X}} \,\dot{\varphi} \Bigl(-2 \,{a}^2 \,\ddot{\varphi} (-2 \,{x} + \,{a} \,\dot{a}) + \,\dot{\varphi} \bigl(3 \,{x}^2 - 6 \,{a} \,{x} \,\dot{a} + \,{a}^2 (-2 \,{a} \,\ddot{a} + 3 \,\dot{a}^2 + 2 \,\dot{x})\bigr)\Bigr)}{\,{a}^4} \,,\end{dmath}\begin{dmath}f_{50}=\frac{6 \,{G_{4,{\phi}{\phi}{\phi}X}} (\,{x} + \,{a} \,\dot{a}) \,\dot{\varphi}^3}{\,{a}^2} + \frac{6 \,{G_{4,{\phi}XXX}} (\,{x} + \,{a} \,\dot{a})^2 \,\dot{\varphi}^4 (- \,{a} \,\ddot{\varphi} + \,\dot{a} \,\dot{\varphi})}{\,{a}^9} -  \frac{6 \,{G_{4,{\phi}{\phi}XX}} (\,{x} + \,{a} \,\dot{a}) \,\dot{\varphi}^3 \bigl(- \,{a}^2 \,\ddot{\varphi} + (\,{x} + 2 \,{a} \,\dot{a}) \,\dot{\varphi}\bigr)}{\,{a}^6} + \frac{\,{G_{4,{\phi}X}} \Bigl(-6 \,{a} \,\ddot{\varphi} (\,{x} + \,{a} \,\dot{a}) (3 \,{x} + \,{a} \,\dot{a}) + 12 \,\dot{\varphi} \bigl(6 \,{x}^2 \,\dot{a} + \,{a} \,{x} (-2 \,{a} \,\ddot{a} + 6 \,\dot{a}^2 - 3 \,\dot{x}) + \,{a}^2 \,\dot{a} (- \,{a} \,\ddot{a} + \,\dot{a}^2 - 2 \,\dot{x})\bigr)\Bigr)}{\,{a}^5} + 6 \,{G_{4,{\phi}{\phi}}} (\frac{\,{x}^2}{\,{a}^2} + \,{a} \,\ddot{a} + \,\dot{x}) + \frac{6 \,{G_{4,{\phi}{\phi}X}} \,\dot{\varphi} \Bigl(3 \,{a}^2 \,\ddot{\varphi} (\,{x} + \,{a} \,\dot{a}) + \,\dot{\varphi} \bigl(-2 \,{x}^2 - 5 \,{a} \,{x} \,\dot{a} + \,{a}^2 (\,{a} \,\ddot{a} - 2 \,\dot{a}^2 + \,\dot{x})\bigr)\Bigr)}{\,{a}^4} -  \frac{6 \,{G_{4,{\phi}XX}} (\,{x} + \,{a} \,\dot{a}) \,\dot{\varphi}^2 \biggl(2 \,{a} \,\ddot{\varphi} (3 \,{x} + 2 \,{a} \,\dot{a}) + \,\dot{\varphi} \Bigl(-9 \,{x} \,\dot{a} + \,{a} \bigl(-5 \,\dot{a}^2 + 2 (\,{a} \,\ddot{a} + \,\dot{x})\bigr)\Bigr)\biggr)}{\,{a}^7} \,,\end{dmath}\begin{dmath}f_{51}=\frac{1}{2}\,x\,f_{7} \,,\end{dmath}\begin{dmath}f_{52}=\frac{1}{2\,x}f_7 \,,\end{dmath}\begin{dmath}f_{53}=\frac{4 \,{G_{4}}}{\,{a}^2} + \frac{4 (1 - 2 \, c) \,{G_{4,X}} \,\dot{\varphi}^2}{\,{a}^4} \,,\end{dmath}\begin{dmath}f_{54}=\frac{1}{2}f_{11} \,,\end{dmath}\begin{dmath}f_{55}=2 \,{G_{4,{\phi}{\phi}X}} \,\dot{\varphi}^2 + \frac{4 \,{G_{4,XXX}} (\,{x} + \,{a} \,\dot{a}) \,\dot{\varphi}^3 (- \,{a} \,\ddot{\varphi} + \,\dot{a} \,\dot{\varphi})}{\,{a}^7} + 6 \,{G_{4,{\phi}X}} \bigl(\,\ddot{\varphi} + \frac{(\,{x} + \,{a} \,\dot{a}) \,\dot{\varphi}}{\,{a}^2}\bigr) + \frac{2 \,{G_{4,{\phi}XX}} \,\dot{\varphi}^2 \bigl(\,{a}^2 \,\ddot{\varphi} -  (2 \,{x} + 3 \,{a} \,\dot{a}) \,\dot{\varphi}\bigr)}{\,{a}^4} + \frac{2 \,{G_{4,X}} \bigl((-3 + \, c) \,{x}^2 + 4 \,{a} \,{x} \,\dot{a} + \,{a}^2 (-2 \,{a} \,\ddot{a} + \,\dot{a}^2 - 4 \,\dot{x})\bigr)}{\,{a}^4} + \frac{\,{G_{4,XX}} \biggl(-4 \,{a}^2 \,\ddot{\varphi} (4 \,{x} + 3 \,{a} \,\dot{a}) \,\dot{\varphi} - 2 \,\dot{\varphi}^2 \Bigl(3 \,{x}^2 - 6 \,{a} \,{x} \,\dot{a} + \,{a}^2 \bigl(-5 \,\dot{a}^2 + 2 (\,{a} \,\ddot{a} + \,\dot{x})\bigr)\Bigr)\biggr)}{\,{a}^6} \,,\end{dmath}\begin{dmath}f_{56}=4 \,{a}^2 \,{G_{4}} \,,\end{dmath}\begin{dmath}f_{57}= \frac{1}{2}f_{11} \,,\end{dmath}\begin{dmath}f_{58}= (\,{x} + \,{a} \,\dot{a})\left(\frac{6 \,{G_{4,X}} (3 \,{x} + \,{a} \,\dot{a})}{\,{a}^4} -  \frac{18 \,{G_{4,{\phi}X}} \,\dot{\varphi}}{\,{a}^2} + \frac{12 \,{G_{4,XX}} (3 \,{x} + 2 \,{a} \,\dot{a}) \,\dot{\varphi}^2}{\,{a}^6} -  \frac{6 \,{G_{4,{\phi}XX}} \,\dot{\varphi}^3}{\,{a}^4} + \frac{6 \,{G_{4,XXX}} (\,{x} + \,{a} \,\dot{a}) \,\dot{\varphi}^4}{\,{a}^8} \right)\,,\end{dmath}\begin{dmath}f_{59}=-12 \,{a}^2 \,{G_{4}} + 12 \,{G_{4,X}} \,\dot{\varphi}^2 \,,\end{dmath}

where the equations of motion for the background fields have been used only in some coefficients such as $f_{51}$. Using these equations one can get rid of second derivatives of $\varphi$ and $a$, and of $x$, but this leads in many cases to longer expressions.\bigskip

\subsection{Coefficients of the quadratic action in its final form}\label{sec appfinalql}

The background functions $m_i$ with $i=1,\,2,\, 3$ relevant to the final form of the quadratic action (\ref{eqn finalql}) are

\begin{dmath}
m_1= 24 \,{a}^4 \,{G_{4}}^2 \Bigl(3 \,{a}^6 \,{G_{4}}^3 (- \,{G_{4,{\phi}}} \,{G_{4,X}} + \,{G_{4,{\phi}X}} \,{G_{4}}) + \,{a}^4 \,{G_{4}}^2 \bigl(\,{G_{4,{\phi}}} \,{G_{4,X}}^2 -  \,{G_{4,{\phi}X}} \,{G_{4,X}} \,{G_{4}} + \,{G_{4}} (- \,{G_{4,{\phi}}} \,{G_{4,XX}} + \,{G_{4,{\phi}XX}} \,{G_{4}})\bigr) \,\dot{\varphi}^2 + \,{a}^2 \,{G_{4}} (- \,{G_{4,{\phi}}} \,{G_{4,X}} + \,{G_{4,{\phi}X}} \,{G_{4}}) (-2 \,{G_{4,X}}^2 + \,{G_{4,XX}} \,{G_{4}}) \,\dot{\varphi}^4 + \bigl(-2 \,{G_{4,{\phi}}} \,{G_{4,X}}^4 + 2 \,{G_{4,{\phi}X}} \,{G_{4,X}}^3 \,{G_{4}} -  \,{G_{4,X}} (4 \,{G_{4,{\phi}X}} \,{G_{4,XX}} + \,{G_{4,{\phi}}} \,{G_{4,XXX}}) \,{G_{4}}^2 + 2 \,{G_{4,X}}^2 \,{G_{4}} (\,{G_{4,{\phi}}} \,{G_{4,XX}} + \,{G_{4,{\phi}XX}} \,{G_{4}}) + \,{G_{4}}^2 (\,{G_{4,{\phi}}} \,{G_{4,XX}}^2 -  \,{G_{4,{\phi}XX}} \,{G_{4,XX}} \,{G_{4}} + \,{G_{4,{\phi}X}} \,{G_{4,XXX}} \,{G_{4}})\bigr) \,\dot{\varphi}^6\Bigr)\,,
\end{dmath}

\begin{dmath}
m_2= -8 \,{a}^4 \,{G_{4,X}}^4 \,{G_{4}} (\,{x} + \,{a} \,\dot{a}) \,\dot{\varphi}^8 \bigl(6 \,{a}^2 \,\ddot{\varphi} + (11 \,{x} + 5 \,{a} \,\dot{a}) \,\dot{\varphi}\bigr) + 4 \,{a}^{12} \,{G_{4}}^5 (\,{x} + \,{a} \,\dot{a}) \bigl(-6 \,{a}^2 \,\ddot{\varphi} + (13 \,{x} + 9 \,{a} \,\dot{a}) \,\dot{\varphi}\bigr) + \,{G_{4}}^2 \Bigl(32 \,{a}^6 \,{G_{4,X}}^3 (\,{x} + \,{a} \,\dot{a}) \,\dot{\varphi}^6 \bigl(3 \,{a}^2 \,\ddot{\varphi} + (2 \,{x} -  \,{a} \,\dot{a}) \,\dot{\varphi}\bigr) + 24 \,{a}^4 \,{G_{4,X}}^2 \,{G_{4,XX}} (\,{x} + \,{a} \,\dot{a}) \,\dot{\varphi}^8 \bigl(\,{a}^2 \,\ddot{\varphi} + (2 \,{x} + \,{a} \,\dot{a}) \,\dot{\varphi}\bigr)\Bigr) + \,{G_{4}}^4 \biggl(8 \,{a}^{14} \,{G_{4,{\phi}{\phi}X}} \,\dot{\varphi}^5 + 8 \,{a}^{11} \,{G_{4,{\phi}XX}} \,\dot{\varphi}^5 (\,{a} \,\ddot{\varphi} -  \,\dot{a} \,\dot{\varphi}) + 8 \,{a}^8 \,{G_{4,XX}} \,\dot{\varphi}^4 \bigl(2 \,{a}^3 \,\ddot{\varphi} \,\dot{a} -  (7 \,{x}^2 + 9 \,{a} \,{x} \,\dot{a} + 4 \,{a}^2 \,\dot{a}^2) \,\dot{\varphi}\bigr) + 8 \,{a}^{10} \,{G_{4,X}} \,\dot{\varphi}^2 \Bigl(\,{a}^2 \,\ddot{\varphi} (5 \,{x} + 3 \,{a} \,\dot{a}) - 2 \,\dot{\varphi} \bigl(\,{x}^2 - 5 \,{a} \,{x} \,\dot{a} + \,{a}^2 (-2 \,\dot{a}^2 + \,\dot{x})\bigr)\Bigr)\biggr) + \,{G_{4}}^3 \biggl(4 \,{a}^4 \,{G_{4,XX}}^2 (\,{x} + \,{a} \,\dot{a}) \,\dot{\varphi}^8 \bigl(2 \,{a}^2 \,\ddot{\varphi} + (\,{x} -  \,{a} \,\dot{a}) \,\dot{\varphi}\bigr) + \,{G_{4,X}} \Bigl(-8 \,{a}^{12} \,{G_{4,{\phi}{\phi}X}} \,\dot{\varphi}^7 + 8 \,{a}^9 \,{G_{4,{\phi}XX}} \,\dot{\varphi}^7 (- \,{a} \,\ddot{\varphi} + \,\dot{a} \,\dot{\varphi}) + 8 \,{a}^6 \,{G_{4,XX}} \,\dot{\varphi}^6 \bigl(- \,{a}^2 \,\ddot{\varphi} (5 \,{x} + 7 \,{a} \,\dot{a}) - 2 \,{x} (\,{x} + 2 \,{a} \,\dot{a}) \,\dot{\varphi}\bigr)\Bigr) + 8 \,{a}^8 \,{G_{4,X}}^2 \,\dot{\varphi}^4 \Bigl(- \,{a}^2 \,\ddot{\varphi} (11 \,{x} + 9 \,{a} \,\dot{a}) + \,\dot{\varphi} \bigl(3 \,{x}^2 + 3 \,{a} \,{x} \,\dot{a} + 2 \,{a}^2 (4 \,\dot{a}^2 + \,\dot{x})\bigr)\Bigr)\biggr)\,,
\end{dmath}
\begin{dmath}
m_3= (\,{x} + \,{a} \,\dot{a})^2 \left(\,{a}^8 \,{G_{4}}^4 \,\dot{\varphi} - 2 \,{a}^4 \,{G_{4,XX}} \,{G_{4}}^3 \,\dot{\varphi}^5 + 4 \,{G_{4,X}}^4 \,\dot{\varphi}^9 - 4 \,{G_{4,X}}^2 \,{G_{4,XX}} \,{G_{4}} \,\dot{\varphi}^9 + \,{G_{4}}^2 \bigl(4 \,{a}^4 \,{G_{4,X}}^2 \,\dot{\varphi}^5 + \,{G_{4,XX}}^2 \,\dot{\varphi}^9\bigr)\right)\,.
\end{dmath}

Using the equations for the background fields one can get rid of second derivatives of $\varphi$, first derivatives of $a$ and of $x$, but this leads in $m_2$ and $m_3$ to much longer expressions.

\section{References}
\bibliographystyle{IEEEtran}
\bibliography{v4HorndeskiCCosmoPert}


\end{document}